\documentclass[a4paper,11pt]{article}

\usepackage{jcappub} 

\usepackage[T1]{fontenc} 

\usepackage{epsfig}
\usepackage{latexsym}
\usepackage{graphicx}
\usepackage{epstopdf}  
\usepackage[justification=RaggedRight]{caption}
\usepackage{amssymb}
\usepackage{subfig}
\usepackage{color}
\usepackage{multirow}
\usepackage{color}
\usepackage{rotating}
\usepackage{ifthen}
\usepackage{epsfig}
\usepackage{booktabs} 
\usepackage{hyperref} 
\usepackage{hyperref}
\hypersetup{
    colorlinks,%
    citecolor=blue,%
    filecolor=blue,%
    linkcolor=blue,%
    urlcolor=blue
}
\def\be{\begin{equation}}
\def\ee{\end{equation}}

\newcommand{\bwt}{\begin{widetext}}
\newcommand{\ewt}{\end{widetext}}
\newcommand{\bdm}{\begin{displaymath}}
\newcommand{\edm}{\end{displaymath}}
\newcommand{\bea}{\begin{eqnarray}}
\newcommand{\eea}{\end{eqnarray}}

\newcommand{\beqra}{\begin{eqnarray}}
\newcommand{\eeqra}{\end{eqnarray}}
\newcommand{\beq}{\begin{equation}}
\newcommand{\eeq}{\end{equation}}

\title{Minimal Decaying Dark Matter and the LHC}

\author{Giorgio Arcadi}

\author{and Laura Covi}

\affiliation{Institute for Theoretical Physics, 
Georg-August University G\"ottingen, 
Friedrich-Hund-Platz~1, G\"ottingen, D-37077 Germany}

\abstract{We consider a minimal Dark Matter model with
just two additional states, a Dark Matter Majorana fermion
and a colored or electroweakly charged scalar, without 
introducing any symmetry to stabilize the DM state. 
We identify the parameter region where an indirect DM signal 
would be within the reach of future observations and the DM relic 
density generated fits the observations.
We find in this way two possible regions in the parameter
space, corresponding to a FIMP/SuperWIMP or a WIMP DM.
We point out the different collider signals of this scenario
and how it will be possible to measure the different
couplings in case of a combined detection.
}

\begin{document} 
\maketitle
\flushbottom

\section{Introduction}

The particle identity of Dark Matter is one of the
most important open questions of astrophysics and particle
physics. In the recent years a strong effort has been
started to pinpoint the Dark Matter candidate from all 
available data, coming from indirect detection, direct 
detection and colliders~\cite{Bertone:2004pz, Kane:2008gb, 
Baer:2009uc}, in particular for WIMP candidates.
We are indeed in a very promising decade, where at the same 
time observations in indirect detection from the FERMI
satellite~\cite{Ackermann:2011wa, Zaharijas:2012dr, Morselli:2012xra} 
and AMS~\cite{Aguilar:2013qda} as well as H.E.S.S. \cite{Abramowski:2012au,Abramowski:2011hh,Abramowski:2010aa} 
and MAGIC \cite{Aleksic:2011jx,Aleksic:2009ir} 
experiments, experiments in direct detection 
like XENON100~\cite{Aprile:2012nq, Aprile:2013doa}, 
and searches for new physics at the LHC detectors 
ATLAS and CMS are very actively pursued.

In these efforts towards a DM detection, decaying Dark Matter 
candidates have been less systematically studied and the
interest has been mostly concentrated on specific well-motivated 
models like decaying sterile neutrinos~\cite{Asaka:2005an,Boyarsky:2005us,Boyarsky:2006zi,Boyarsky:2006fg,Herder:2009im}, 
the gravitino with R-parity breaking
\cite{Takayama:2000uz,Buchmuller:2007ui,Lola:2007rw,Covi:2008jy,Bomark:2009zm,Buchmuller:2009xv, Bobrovskyi:2010ps,Bajc:2010qj,Bobrovskyi:2011vx, 
Vertongen:2011mu} or leptophilic decaying Dark Matter, in order 
to explain the PAMELA excess~\cite{Kyae:2009jt, Ibarra:2009bm, Garny:2010eg}, or
 more recent results form AMS-02~\cite{Ibe:2013nka,Jin:2013nta}. 
In this paper we would like to extend these studies in a 
more comprehensive way in a simplified scenario, where the 
Dark Matter sector includes a neutral Majorana spinor, 
our Dark Matter candidate, and a single heavy colored or 
electroweakly charged scalar, possibly in a mass range 
within the reach of the LHC collider.
We will restrict us here to renormalizable couplings between
the two fields and the Standard Model since such case is
complementary to the ``non-renormalizably'' coupled Dark Matter
candidates like the gravitino with R-parity violation~\cite{Buchmuller:2007ui,Ji:2008cq,Restrepo:2011rj,Arcadi:2011yw}.
Along the majority of this paper we will also refer to a rather 
simple realization in which all the relevant interactions, 
can be expressed in terms of two additional Yukawa-type couplings. 

This setting gives us a minimal model, which can be embedded 
in and describe the low-energy limit of more complicate and 
richer models, like for example supersymmetric models with 
R-parity violation.  In fact such a simplified scenario can still reflect 
the key features of those models as long as the rest of the 
states and other couplings are outside the reach of the LHC.

Our hope is to individuate the parameter space where the
model is both cosmologically viable and observable through 
multiple signals and see if future searches can identify the 
model from the data. In particular we are interested in the possibility 
of a concurrent detection of such decaying 
dark matter both in indirect detection and at the LHC and in how
the two datasets could be best used to reconstruct the model.

Regarding the Dark Matter generation, we will see that it
can be accomplished in this simple setting by different 
mechanisms, either the {\it FIMP/freeze-in} mechanism~\cite{McDonald:2001vt, 
Covi:2002vw, Asaka:2005cn, Hall:2009bx,Cheung:2010gj, Kang:2010ha, Frigerio:2011in} or the non-thermal production from 
the decay of the scalar field a' la {\it SuperWIMP}~\cite{Covi:1999ty,Feng:2003xh,Feng:2003uy} 
or even the {\it freeze-out/WIMP} mechanism. 
The latter case has already been analyzed
in an equivalent~\footnote{In the referred paper the DM 
particle is assumed stable, while in our model an additional
coupling allows also for its decay. But in the WIMP regime
this second coupling has to be very strongly suppressed
and therefore the dominant interactions coincide with
\cite{Garny:2012eb}.} model by \cite{Garny:2012eb}, so we 
will not repeat their analysis here, but concentrate on 
different parts of the parameter space. 

After determining the couplings and mass ranges allowing 
for the right Dark Matter density, we will discuss on general 
grounds, whether these regions can be searched at the LHC and 
which couplings are accessible there. 
We will, in particular, highlight the scenarios in which both 
couplings can be measured combining informations from indirect 
detection and collider. The detailed study of the signals possibly 
associated to the scenarios that will emerge, will be instead 
 object of a dedicated paper.
 
This paper is organized as follows: First we define our minimal
model in Section 2, then we discuss the indirect detection
signatures due to the decay of the DM particle in Section 3.
We describe the cosmology of the scenario depending on
the couplings and single out the region giving the right
abundance for Dark Matter in Section 4. Section 5 shortly
discusses other non cosmological constraints on the scenario,
while section 6 is dedicated to the LHC signals.
Finally we discuss our results and conclude in section 7.

\section{Definition of the model}

We consider a minimal model in which the dark matter is coupled 
to a standard model fermion and a scalar field $\Sigma_f$, charged 
under at least part of the Standard Model gauge group, through 
a Yukawa-type interaction whose general form is given by
 \cite{Garny:2010eg,Garny:2011ii,Garny:2012vt}: 
\begin{equation}
L_{\rm eff} =\lambda_{\psi f L} \bar{\psi} f_L \Sigma_f^{\dagger} 
+\lambda_{\psi f R}\bar{\psi}  f_R \Sigma_f^{\dagger} +h.c.
\label{eq:DMint}
\end{equation}
taking into account the possibility of different couplings with fermions 
of different chiralities. The DM is assumed to be a Majorana fermion SM singlet. 

According a minimality principle only one $\Sigma_f$ field is 
introduced. As a consequence the DM will couple only with quarks or leptons 
depending on whether $\Sigma_f$ carries color or only electroweak charge. 
These two cases will be respectively referred to, for simplicity, as hadronic and 
leptonic models. In addition only one among the couplings 
$\lambda_{\psi f R}$ and $\lambda_{\psi f L}$ will be present, according 
the SU(2) quantum numbers of $\Sigma_f$. 

With the aim of proposing a model with a rich collider phenomenology, it may
seem more promising to concentrate on hadronic models because of their direct 
coupling of the DM with quarks. However, as will be shown later on, leptonic models offer 
very interesting possibilities, even in virtue of their weaker limits.
Moreover the next future energy upgrade of the LHC will likely extend its reach to 
electroweak processes and for this reason we consider both types of models 
in our study.
 
Notice that the scalar field $\Sigma_f$ can couple directly to SM fermions in an 
analogous manner to sfermions in RPV supersymmetry~\cite{Barbier:2004ez}.
Within the hadronic models, we distinguish three configurations:
\begin{itemize}
\item The scalar field is an SU(2) singlet with electromagnetic charge equal 
to -1/3, thus equivalent to the squark $\tilde{d}_R$ and then, in analogy, 
will be referred as $\Sigma_d$. Then the following two operators with 
SM field are allowed:
\begin{equation}
L_{\rm eff}=\lambda^{'}_{d} \bar{l}^c_{R}q_L \Sigma_d^{\dagger}+
\lambda^{''}_{d} \bar{u}_R d^c_L \Sigma_d^{\dagger}+h.c.
\end{equation}
 \item  The scalar has the quantum numbers of the squark $\tilde{u}_R$, leading 
only to the operator:
 \begin{equation}
L_{\rm eff}=\lambda^{''}_{u} \bar{d}_R d^c_L \Sigma_u^{\dagger}+h.c.
 \end{equation}
\item The last possibility is that we have an $SU(2)_L$ doublet:
\begin{equation}
\Sigma_q \equiv \left(
\begin{array}{c}
\sigma_u \\
\sigma_d
\end{array}
\right)
\end{equation}
giving:
\begin{equation}
L_{\rm eff}=\lambda^{'}_{q}  \bar{d}_R l_L \Sigma_q+h.c.
\end{equation}
\end{itemize}  
In the leptonic models instead, the scalar field is charged only under
EW interactions and couples to leptons according the following two operators:
\begin{equation}
L_{\rm eff}= \lambda_{e} \bar{\nu}^c_L l_L \Sigma_e^{\dagger}+h.c.
\end{equation}
\begin{equation}
L_{\rm eff}={\lambda}_{l} \bar{e}_R l_L \Sigma_l+ \lambda_{l}^{'} \bar{d} _R q_L \Sigma_l+h.c.
\end{equation}
in which the scalar field is, respectively, an $SU(2)$ singlet or doublet.

For simplicity we assume all the couplings between $\Sigma_f$ and the SM fermions to 
be flavor universal mentioning, where relevant, possible effects when this last 
assumption is relaxed.

\section{Dark Matter Indirect Detection signatures}

The effective Lagrangians introduced above induce DM three-body decays into three SM fermions.  
Unless explicitly stated final state particles will be assumed massless. In this case the 
DM decay rate simplifies to \cite{Garny:2010eg,Garny:2012vt}:
\begin{equation}
\Gamma_{\rm DM}=\frac{ |\lambda_{\psi f}|^2 |\lambda^{'}|^2}{128 {\left(2 \pi\right)}^3} 
\frac{m_\psi^5}{m_{\Sigma_f}^4}
\label{eq:Gamma_DM}
\end{equation}
where we have generically indicated with $\lambda'$ the coupling between 
$\Sigma_f$ and SM fermions only. We are giving here the decay into a single 
generation, in principle it can be easily generalized to more flavors by 
taking into account the flavor index of the couplings and substituting 
$ |\lambda_{\psi f} |^2 = \sum_i |\lambda_{\psi f i}|^2 $, and 
$ |\lambda^{'}|^2 \rightarrow \sum_{ij} \lambda^{'}_{ij}\lambda^{'\,\dagger}_{ji} $.

In addition two-body decays into gauge bosons (W,Z and photons) and 
leptons/neutrinos may arise at loop level. 
In such a case the branching ratios of these processes are suppressed at least by 
a factor:
\begin{equation}
\frac{9 \alpha_{\rm em}}{8 \pi} \frac{\sum_j{\left(\sum_i Q_i \lambda_{\psi q\, i} \lambda_{ij}^{'}\right)}^2}{ |\lambda_{\psi q}|^2 |\lambda^{'}|^2}
\end{equation}
where $Q_i$ is the charge of the quark in units of $e$, for $\Sigma_{f}$ 
interacting with quarks, and
\begin{equation}
\frac{3 \alpha_{\rm em}}{8 \pi} \frac{\sum_j{\left(\sum_i \lambda_{\psi l\,i} \lambda_{ij}^{'}\right)}^2}{ |\lambda_{\psi l}|^2 |\lambda^{'}|^2}
\end{equation}
for the case of interactions with leptons. Further suppression may occur when 
the relevant diagrams require a chirality flip. Nonetheless, especially in the 
case of decay into $\gamma$-rays, constraints competitive with three-body decays 
can be obtained \cite{Garny:2010eg,Garny:2012vt}. In the following we will 
briefly illustrate, for each of the models introduced above, the most relevant 
processes and the search channels with the highest experimental sensitivity.

For hadronic models, i.e. $\Sigma_{d,u,q}$, the most relevant search channel 
is the one into antiprotons. Constraints for three-body decays of the type 
$q \bar{q} \nu$, for all different quark flavors, have recently been derived 
in \cite{Garny:2012vt}. 
The authors report limits ranging between $10^{26}\mbox{s}$ and $10^{28}\mbox{s}$ 
as consequence of the uncertainties in the cosmic rays diffusion parameters and 
in the treatment of the background. Note that the limits are slightly stronger
for heavier quarks, especially for the case of the top final state, but such
channel opens up only for heavy Dark Matter with mass above $2 m_{t} \sim 345 $ GeV.
Unless differently stated we will adopt here a reference limit on the DM lifetime 
of $10^{27}$ s for each single channel, which we will assume to hold as well 
for the other possible hadronic decay channels, namely $u \bar{d}\bar{d}$ and 
$u \bar{d} l$. 

Radiative decays instead do not play an important role. Indeed, in the case that 
$\Sigma_f$ behaves like a $\tilde{d}_{R}$ squark, loop decays require a 
chirality flip on an internal fermion line resulting suppressed at least 
by one power of the bottom mass with the exception of the decay into a $W$ and 
a lepton which can allow for the mass insertion of a top. 
The case of $\Sigma_u \equiv \tilde{u}_{R}$ also gives no relevant one-loop 
decays. A potentially interesting case is instead the one in which 
the scalar $\Sigma_f$ is an $SU(2)$ doublet, $\Sigma_q $. Indeed it features 
the decay into a Higgs boson and a neutrino whose rate is suppressed only 
by a loop factor and, eventually, by a third generation Yukawa. 
For Higgs masses in agreement with the recent LHC measurement, the dominant 
signal would be $b \bar{b} \nu$ which is already taken into account in the 
limit for the general $q \bar{q} \nu$ discussed above. 
Peculiar signatures could be obtained in the case of Higgs decays into 
$\gamma$-rays \cite{Ibarra:2012dw}, however the potential signal results 
suppressed by the branching ratio of the Higgs into $\gamma$-rays which 
is of the order of $10^{-3}$. 

Moving to leptonic models, in both cases we have the three level decay into 
the combination $l\bar{l}\nu$. If the scalar field is an $SU(2)$ doublet, i.e.
a' la $\Sigma_{\ell}$, the decay channels $\bar{u}dl, u\bar{d}\bar{l}$ and 
$d\bar{d}\nu$ are present as well. If decays into leptons only are dominant, 
the greatest experimental impact has been found to be on the $\gamma$-ray 
continuum, overcoming even electron-positron searches. 
In this work we take as reference the limits derived by \cite{Cirelli:2012ut}. 
For the case $\Sigma_{\ell}$, the loop induced decay into $\gamma\,\nu$ 
can provide complementary constraints from searches of monochromatic $\gamma$-lines.  

\begin{figure}[htb]
 \subfloat{\includegraphics[width=6.5cm, height=6.5cm, angle=360]{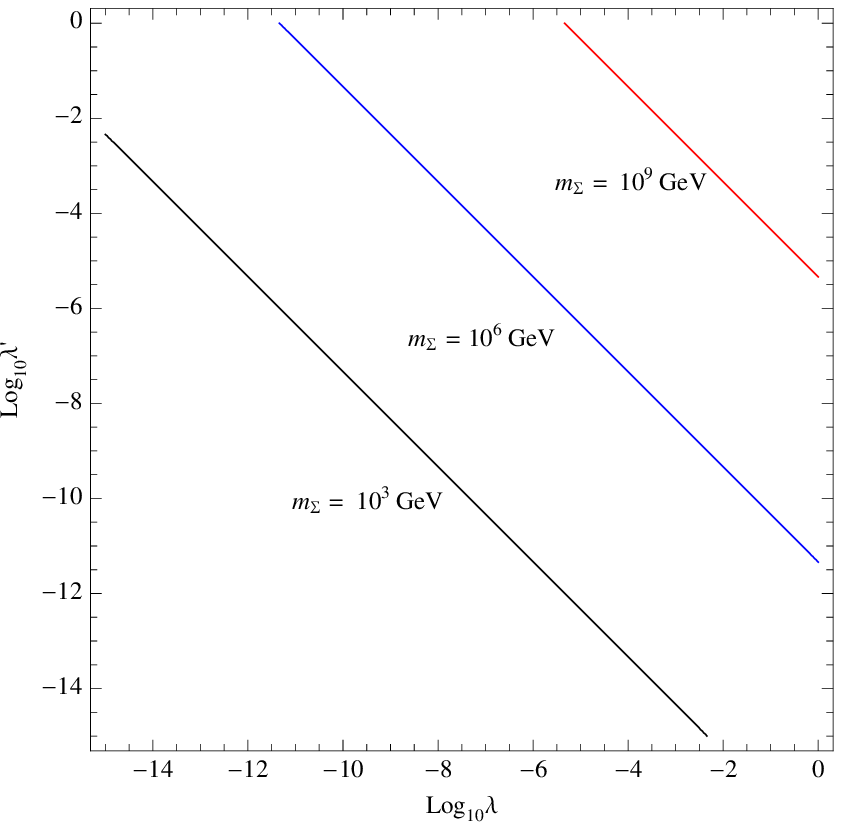}}
 \hspace{5mm}
\subfloat{\includegraphics[width=6.5cm, height=6.5cm, angle=360]{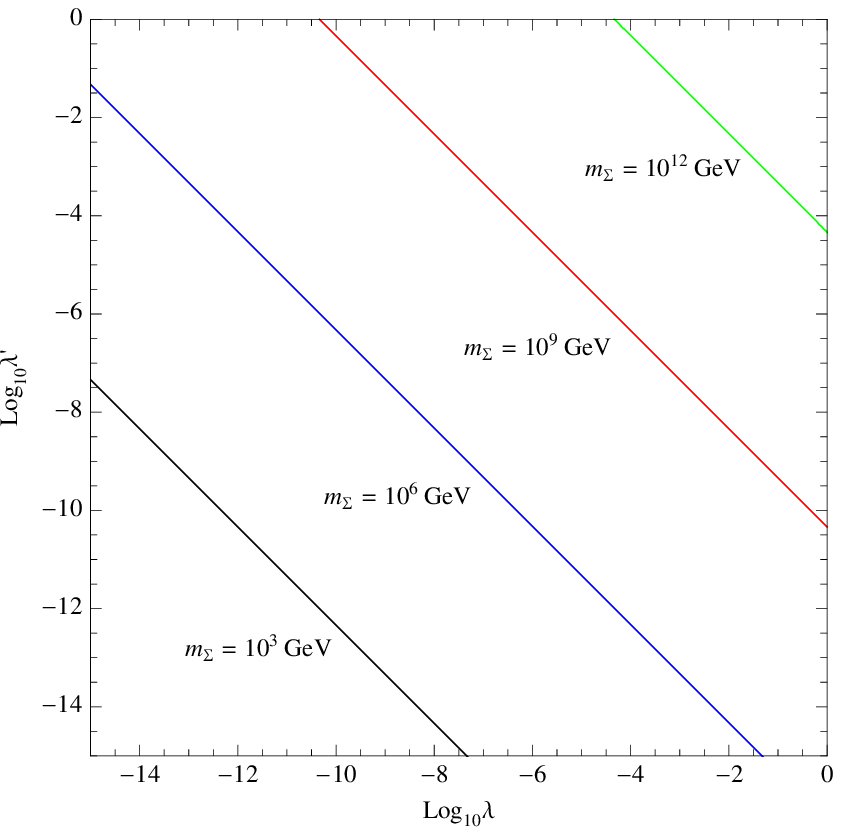}}
\caption{Contours of $\tau_{\rm DM}=10^{27}\mbox{s}$ for the values of $m_{\Sigma_f}$ reported in the plot. 
In the plots the DM mass have been fixed, from left to right, to 1 and 100 GeV.}
\label{fig:Iplot}
\end{figure}

Unless differently stated, along this work we will focus the attention on the 
tree-level decays. From fig.~(\ref{fig:Iplot}) we can get a first insight 
of the impact of the limits from cosmic rays on the couplings of our model.

As it is evident, the two couplings (or at least one of the two), must be highly 
suppressed unless the mass of the scalar is lifted above $10^{12}\mbox{GeV}$
or even close to the GUT scale for moderate values of the DM mass 
(this kind of scenario is discussed, for example 
in~\cite{Eichler:1989br,Arvanitaki:2008hq,Hamaguchi:2008ta}). 
This possibility is however not very interesting in the context of a collider
detection since, in order to obtain a richer phenomenology at the LHC,
the mass of the scalar field $\Sigma_f$ should not exceed the value of 
few TeV's. Therefore, in the following we will consider values of the 
product of couplings $\lambda_{\psi\,f}\,\lambda^{'}$ of the order 
of $10^{-(16 \div 22)}$, depending on the DM mass, consistent with
a SM charged scalar field within the LHC reach. 

The potential correlation between DM Indirect Detection and collider signals
is due to the fact that the same couplings $\lambda$ and $\lambda'$, involved 
in the DM decay rate, induce decays of $\Sigma_f$ which can be observed at the 
LHC as well as, in the case of DM couplings with the quarks, direct production 
of dark matter. Leaving this second possibility to further discussion, we will 
start examining the impact of the ID limits on a possible detection of the 
scalar particle $\Sigma_f$.   

Assuming that $m_{\Sigma_f}$ is the only relevant mass scale, its decay rate for 
a particular channel is given by:
\begin{equation}
\label{eq:Sigma_rate}
\Gamma=\frac{\tilde{\lambda}^2}{8 \pi} m_{\Sigma_f}\,\,\,\,\,\mbox{with}\,\,\,\tilde{\lambda}=\lambda_{f\psi\,L,R}, \lambda_{f}, \lambda^{'}_{f}, 
\lambda^{''}_{f} 
\end{equation}

We can then express the rate (or equivalently the lifetime) of a given process 
of the type $\psi \rightarrow f \bar{f}^{'} f^{''}$ in term of the rates of the 
decays $\Sigma_f \rightarrow f \psi$ and $\Sigma_f \rightarrow \bar{f}^{'} f^{''}$ or, 
more relevant for collider purposes, in terms of the decay lengths\footnote{We neglect for simplicity eventual boost factors.} 
$l = c \tau$. 

\begin{figure}[htb]
\begin{center}
\subfloat{\includegraphics[width=9cm, height=7 cm, angle=360]{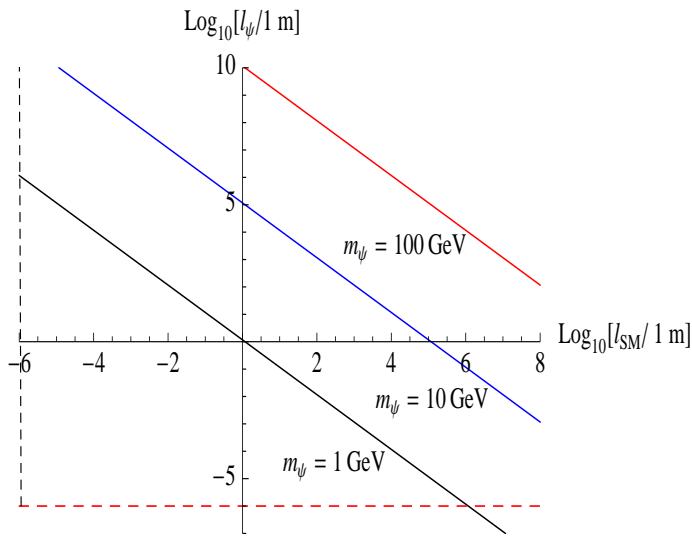}}
\end{center}
\caption{Value of the $\Sigma_f$ decay length into DM and one fermion versus 
the decay length into SM states only, as imposed by eq.(\ref{eq:corr1}), assuming the DM 
lifetime to be $10^{27}$ s. In addition $m_{\Sigma_f}$ is fixed at 1 TeV while for $m_\psi$ we 
have chosen three sample values reported on the plot. For future utility we have indicate 
the value $100\mu\mbox{m}$ below which we can observe at LHC prompt decay of the 
field (see next sections).}
\label{fig:Ipsi}
\end{figure}

We can then relate the DM lifetime directly to the decay lengths of 
$\Sigma_f$ in the two channels, namely DM + SM fermion and SM fermions only,
and obtain:
\begin{align}
&l_{\Sigma, DM}=  \frac{ c \hbar}{\Gamma(\Sigma_f \rightarrow f\,\psi)}
=\frac{\tau_{\psi} c^2 \hbar}{16 \pi} \frac{m_{\Sigma_f}^6}{m_\psi^5} l_{\Sigma,SM}^{-1}\nonumber\\
& \simeq 1.17\,\mbox{m} {\left(\frac{m_{\Sigma_f}}{1\mbox{TeV}}\right)}^{-6} 
{\left(\frac{m_\psi}{1\mbox{GeV}}\right)}^{5} 
{\left(\frac{l_{\Sigma,SM}}{1\,\mbox{m}}\right)}^{-1}\left(\frac{\tau_\psi}{{10}^{27}\mbox{s}}\right)\; .
\label{eq:corr1}
\end{align} 
We see therefore that for a DM lifetime at the edge of detection, $\Sigma_f$ decay
lenghts within the LHC detectors and of a similar order of magnitude can be achieved.
Depending on the masses of the Dark Matter and of the scalar field $\Sigma_f$, 
it may therefore be possible to realize a scenario with a contemporary detection of
both $\Sigma_f$ decays and with dark matter lifetime observable in the next future.
We plot in fig.~(\ref{fig:Ipsi}) the relation given by eq.~(\ref{eq:corr1}) 
for some sample values of the DM mass and assuming a reference value 
$m_{\Sigma_f}=1 \mbox{TeV}$ and, as already stated, a DM lifetime of $10^{27}$ s.

In the next section we will investigate whether this kind of scenario is still 
viable when the cosmological constraints on the DM abundance are taken into account.  

\section{Cosmology}

In this section we will examine the mechanisms which allow to generate the
right Dark Matter density from the same couplings considered in section 2
and discuss how the DM density constraints reduce the parameter space of the
models.
As will be evident, leptonic and hadronic models share most of the main cosmological features. 
For this reason we will adopt a simplified setup in which the interactions are 
encoded in two generic couplings $\lambda$ and $\lambda^{'}$ representing,
respectively, the interactions of the scalar field $\Sigma_f$ with the dark matter 
and the SM states only. For the latter we will explicitly mention when possible 
differences between the hadronic and leptonic models may arise. 
No assumption on the size of these couplings is made for the moment, but 
they will be determined only from the requirement of DM viability and, 
in the next sections, from detection limits. 

On general grounds, the DM abundance in the primordial plasma can be computed 
by solving a system of coupled Boltzmann equations for the $\Sigma_f$ and $\psi$ field. 
On the other hand the scalar field $\Sigma_f$ also feels gauge interactions 
with ordinary matter which guarantee efficient annihilation and scattering
processes keeping it in thermal equilibrium until it freezes-out. 
The evolution of number densities of the two particles can then be decoupled 
and the main trends can be identified on purely analytical grounds. 
In the following we will anyway present a numerical treatment of the 
$\psi$ and $\Sigma_f$ Boltzmann equations to confirm the validity of this assumption.

Depending on the value of the coupling $\lambda$, different generation mechanisms 
may account for the DM relic density. Two main scenarios can be identified.
The first is characterized by very low values of $\lambda$, below $10^{-7}$ 
as will see in the next section. For such low value of the coupling, the DM cannot
be in thermal equilibrium in the Early Universe. Two generation mechanisms 
are nevertheless active in this case~\footnote{We will neglect here other
possible non-thermal mechanisms like production via inflation decay or
during preheating/reheating.}, both involving the decay of $\Sigma_f$
into Dark Matter:
 \begin{itemize}
\item Dark matter is produced by the decays (and in principle also
scatterings, as we discuss in the Appendix) of the scalar $\Sigma_f$ while 
it is still in thermal equilibrium. This generation process is dubbed 
FIMP~\cite{Hall:2009bx,Cheung:2010gj}.
\item Non-thermal production through the decay of the scalar field
after is has undergone freeze-out, as in the SuperWIMP 
case~\cite{Covi:1999ty,Feng:2003xh,Feng:2003uy} . 
\end{itemize}

The second option is instead to have $\lambda$ of order one and 
the DM in thermal equilibrium in the early stages 
of the cosmological history and produced according the freeze-out paradigm. 
As can be argued from eq.~(\ref{eq:Gamma_DM}), in this last case ID limits 
require an extreme suppression of the coupling $\lambda^{'}$; as a consequence 
this second case essentially coincides with standard WIMP models which have 
already been object of numerous studies across the literature. The only
additional feature would be to have at the same time an annihilation and
a decay signal in Indirect Detection, at the cost of a quite strong fine-tuning.
Nonetheless we will briefly reexamine this scenario in light of the most recent 
experimental updates showing how well such WIMP DM compares with LHC,
especially in case of DM couplings with quarks. 

In the next subsections we will investigate in more detail the cosmology of these two scenarios \footnote{Notice that an intermediate scenario exists as well in which the
DM is initially in thermal equilibrium and decouples while it is relativistic or semi-relativistic \cite{Drees:2009bi}. However the correct relic density
is achieved for extremely low values of the DM mass such that it cannot be regarded as a cold or even warm DM candidate.}.

\subsection{FIMP/Super WIMP regime}

As already stated, the first case that we are going to investigate is the one in 
which the dark matter coupling with ordinary matter is too weak to allow for 
thermal equilibrium in the whole cosmological history.
We can approximately determine the range of values of $\lambda$ for which DM
cannot reach thermal equilibrium by imposing that the ratio of the rate of decay 
of $\Sigma_f$ into DM particles over the Hubble expansion rate is lower than 
one at temperatures of the order of $m_{\Sigma_f}$. 
From this requirement we get:
\begin{equation}
 \lambda^2 < 8 \pi \sqrt{g_{*}} 1.66 \frac{m_{\Sigma_f}}{M_{\rm pl}} g_{\Sigma}^{-1}
 \end{equation}
For masses of the scalar field of within the reach of the LHC, namely below 3 TeV, 
we have then that $\lambda \lesssim 10^{-7}$.

In such case, starting with zero number density, a Dark Matter abundance arises 
from the combination of two mechanisms. 
The first is the mechanism dubbed {\it freeze-in} in which the DM is produced 
by the decays of the scalar field $\Sigma_f$ while it is still in thermal equilibrium. 
The DM relic density is then tightly related to the decay rate of the latter into 
the DM itself as \cite{Hall:2009bx}:
\begin{equation}
\Omega^{FI}h^2 = \frac{1.09 \times 10^{27} g_\Sigma}{g_{*}^{3/2}} \frac{m_\psi \Gamma(\Sigma_f \rightarrow \psi f)}{m_{\Sigma_f}^2}
\end{equation}
where $g_\Sigma$ are the number of internal degrees of freedom of $\Sigma_f$. 
By using (\ref{eq:Sigma_rate}) the expression above can be re-expressed as:  
\begin{equation}
\label{eq:Freezein}
\Omega^{\rm FI} h^2 =\frac{1.09\times 10^{27} g_{\Sigma}}{g_{*}^{3/2}}\frac{\lambda^2 x}{8\pi}
\end{equation}
where $x=m_\psi/m_{\Sigma_f}$ and, for simplicity, we have again assumed the scalar field to be much heavier than 
the DM candidate in order to neglect kinematical suppression factors in the decay rate.

The second mechanism is the SuperWIMP mechanism, producing DM from the decay 
of the scalar field after it has undergone chemical freeze-out.
The contribution from this production mechanism can be expressed as:
\begin{equation}
\Omega_\psi^{SW}h^2 = x\,BR(\Sigma_f \rightarrow \psi f) \Omega_\Sigma h^2
\label{eq:SuperWIMP}
\end{equation}
where $\Omega_\Sigma h^2$ represents the relic density of $\Sigma_f$ computed as if 
it were stable. Notice that in our case the relic density is proportional to the 
branching ratio of $\Sigma_f$ into DM, since we have more than one decay channel present.  
This quantity could in principle be measured at the LHC if both the $ \Sigma_f $ 
decays are accessible.

These two contributions to the DM relic density are very different: indeed freeze-in 
is only sensitive to the interactions between the DM and the scalar field, mediated 
by $\lambda$, while the SuperWIMP contribution depends also on the gauge interactions 
determining the relic density of the scalar field, and on both couplings 
$\lambda$ and $\lambda'$ as encoded in the branching ratio.  Due to this feature, 
several possible scenarios may appear depending on the quantum numbers of $\Sigma_f$.
 On general grounds, we expect $\Omega_{\Sigma}$ to be very low for a charged relic, 
as consequence of the efficient interactions of the scalar field, thus suppressing 
the SuperWIMP contribution at low scalar masses. 
In order to address more precisely this issue we will now estimate $\Omega^{SW}$ for 
some realizations of the hadronic and leptonic models.
 
\begin{figure}[htb]
 \begin{center}
\subfloat{\includegraphics[width=6cm, height=6cm, angle=360]{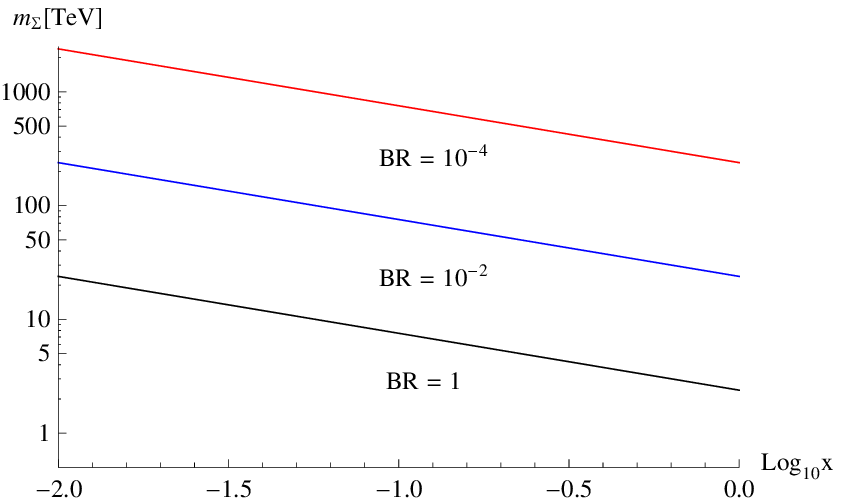}}
\hspace{1cm}
\subfloat{\includegraphics[width=6cm, height=6cm, angle=360]{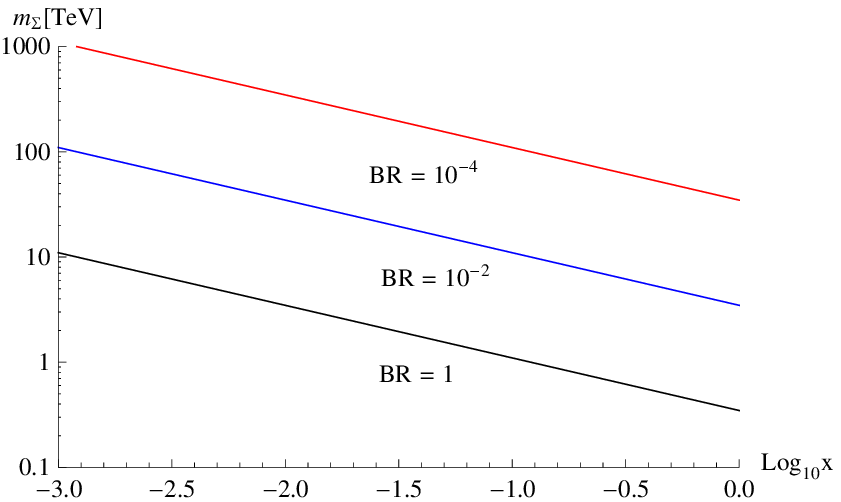}}
 \end{center}
\caption{Values of $m_{\Sigma_f}$, as function of $x$, for which the correct DM relic density is achieved through the SuperWIMP mechanism, for some fixed 
values of $BR(\Sigma_f \rightarrow \psi q)$ reported in the plot. In the left panel the scalar $\Sigma_f$ is a colored state singlet under $SU(2)$ while in 
the right panel it is charged only under hypercharge.}
\label{fig:BRWS}
\end{figure}

Let's consider first the case of color charged SU(2) singlet $\Sigma_d$. 
Adopting the velocity expansion defined in \cite{Berger:2008ti} we have, 
at the leading order, for the annihilation into gluons:
\begin{equation}
\langle \sigma v \rangle \rightarrow \frac{14}{27} \frac{\pi \alpha_s^2}{m_{\Sigma_d}^2}
\end{equation}
Plugging in this result we obtain:
\begin{equation}
\label{eq:sigma_relic_an}
\Omega_{\Sigma}h^2 \approx 7.72 \times 10^{-4} {\left(\frac{m_{\Sigma_d}}{1 \mbox{TeV}}\right)}^2  {\left(\frac{\bar{g}_{\rm eff}^{1/2}}{10}\right)}^{-1} \frac{m_{\Sigma_d}}{T_{\rm f, \Sigma_d}}
\end{equation}
where $T_{\rm f, \Sigma}$ is the freeze-out temperature of the field $\Sigma_d$. 
Notice that the we have included a factor 2 in the relic density in order to take 
into account the contribution of the charge conjugate state of $\Sigma_d$.    
The DM relic density is thus given by:
\begin{equation}
\label{eq:SuperWIMPC}
\Omega^{\rm SW} h^2 = 7.72 \times 10^{-4} {\left(\frac{m_{\Sigma_d}}{1 \mbox{TeV}}\right)}^2 \frac{m_{\Sigma_d}}{T_{\rm f,\Sigma_d}} x\, BR\; .
\end{equation}
Assuming the branching ratio $BR$ to be a freely varying parameter, we have computed, 
as function of $x$, the values of $m_{\Sigma_d}$, for which $\Omega^{\rm SW}$ is of 
the order of the cosmological value of the DM relic density, i.e.
 $ \Omega_{CDM} h^2 = 0.11$ \cite{Hinshaw:2012fq,Ade:2013zuv}, 
(see also fig.~(\ref{fig:BRWS})), obtaining a minimal value of 
around 2 TeV for $x \sim 1$ and $BR \sim 1$ for $T_{\rm f,\Sigma_d} = m_{\Sigma_d}/30$.

Analogous limits can be derived for the other possible configurations of the field 
$\Sigma_f$ depending on the relevant cross section. 
If $\Sigma_f$ interacts only through hypercharge, its pair annihilation 
cross section into gauge bosons is given by:   
\begin{equation}
\langle \sigma v \rangle \simeq \frac{3 \pi \alpha_{\rm em}^2}{16 m^2_{\Sigma_e}}
\end{equation}
Moving to the case of the $SU(2)$ doublet, the relic density can be computed analogously 
to the previous cases adopting the cross section:
\begin{equation}
 \langle \sigma v \rangle \simeq \frac{2 \pi \alpha^2_{\rm em}}{m^2_{\Sigma_l}}
\end{equation}
where we have taken degenerate masses $m_{\Sigma_l}$.
Then the SuperWIMP contribution is of the order of the cosmological DM density 
for values of the mass of the scalar of around 1.5 TeV.

The DM relic density can be estimated as the sum of 
(\ref{eq:Freezein}) and (\ref{eq:SuperWIMP}). In order to validate this result 
we have also employed a more systematic approach by solving, through a numerical code, 
the coupled Boltzmann equations for the DM and the scalar field. In particular 
we have taken into account possible additional contributions to freeze-in 
from $2 \rightarrow 2$ scatterings of the DM particles with scalar field $\Sigma_f$ 
and other thermal bath states. We found that in this case indeed these additional
terms are negligible and the formulas (\ref{eq:Freezein}) and (\ref{eq:SuperWIMP})
well-describe the numerical results. All the details of our derivation are summarized 
in the appendix. For definiteness we have focused on a specific case of the 
$SU(2)$ singlet $\Sigma_e$. Fig.~{(\ref{fig:numerics}) reports some sample outcomes 
of our numerical analysis.
\begin{figure}[htb]
\subfloat{\includegraphics[width=7.5 cm, height=6.5 cm, angle=360]{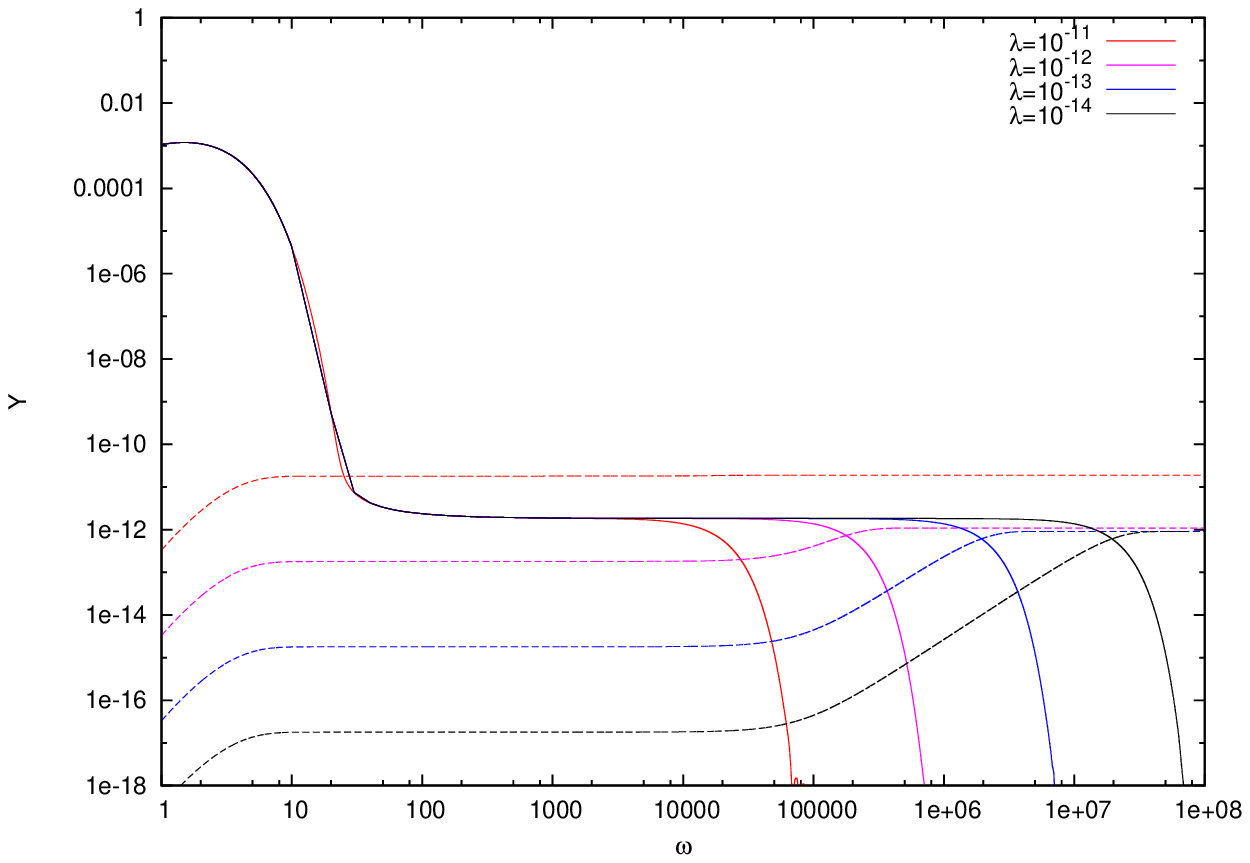}}
\hspace{5 mm}
\subfloat{\includegraphics[width=7.5 cm, height=6.5 cm, angle=360]{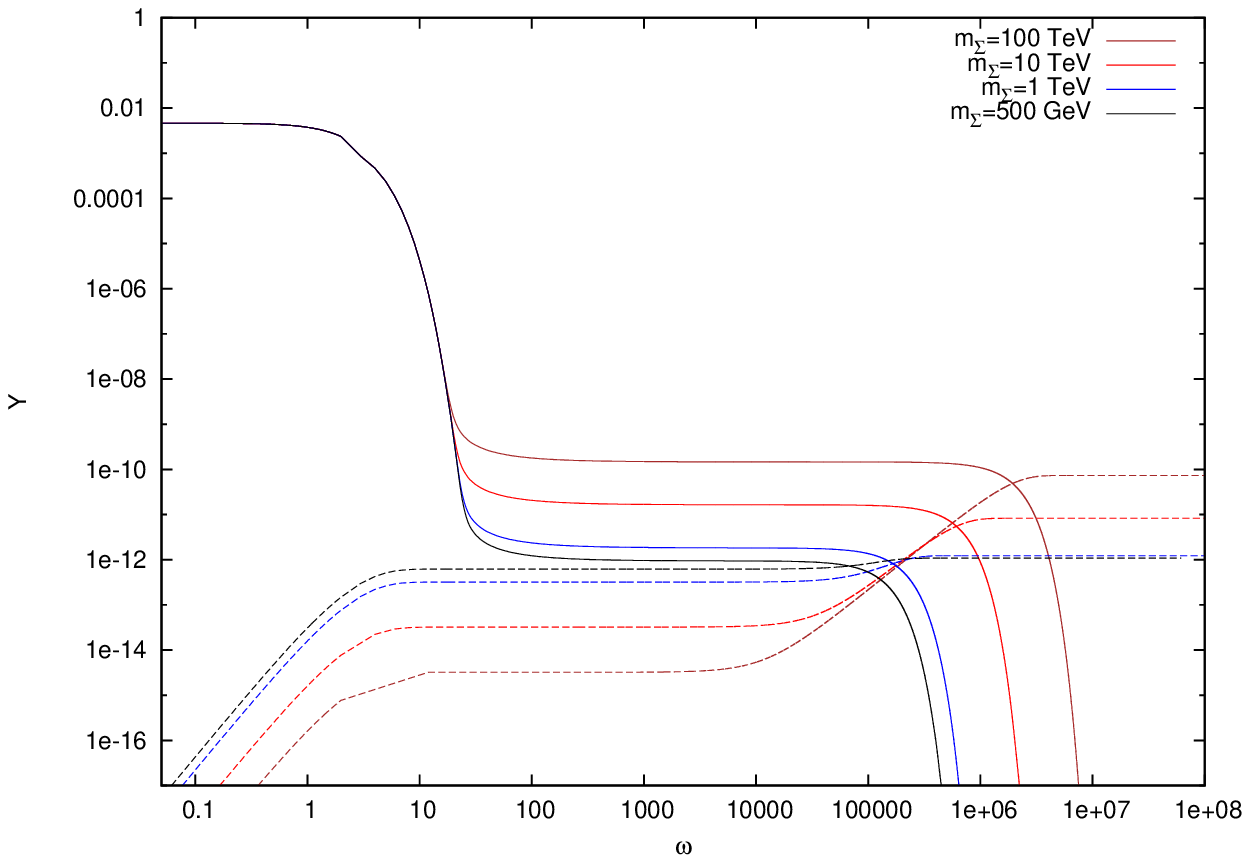}}
\caption{Left panel: Dark matter and $\Sigma_e$ yields, represented respectively with dashed and solid lines, as function of $\omega=m_{\Sigma_e}/T$ 
for a model with $m_\psi=100\mbox{GeV}$ and $m_{\Sigma_e}=1 \mbox{TeV}$. The effective couplings $\lambda$ and $\lambda'$ have been assumed 
equal and assigned to the values reported in the plot. Right panel: Yield of the DM and $\Sigma_e$ for $m_\psi=100\mbox{GeV}$ 
and different values of $m_{\Sigma_e}$. The couplings $\lambda$ and $\lambda'$ have been both assumed equal to $10^{-12}$.}
\label{fig:numerics}
\end{figure} 

In the first plot we give the DM and scalar yields, as function of $\omega=m_{\Sigma_e}/T$, 
for a fixed value of the pair $(x,m_{\Sigma_e})$, namely $(0.1, 1\, \mbox{TeV})$, 
while allowing $\lambda$ to vary. The other coupling $\lambda^{'}$ has been assumed 
equal to $\lambda$ in such a way that $BR(\Sigma_e \rightarrow l \psi) = 1/2$, 
thus allowing for sizable contributions from both DM production mechanism. 
The range of variation of $\lambda$ has been chosen so that the DM relic density is 
within one order of magnitude of the cosmological value. As already expected the scalar 
field undergoes ordinary freeze-out and subsequently out-of-equilibrium decay. 
Regarding the behavior of the DM abundance we notice the freeze-in occurring, 
as expected, while the scalar field is still in thermal equilibrium and being 
the dominant mechanism for values of $\lambda$ of the order of $10^{-11}$. 
By decreasing $\lambda$ the FIMP contribution decreases in favor of the 
SuperWIMP one. This last mechanism entirely determines the DM relic density for 
the lowest values of $\lambda$. As evident from the plot, in this last case the 
DM yield is independent from the lifetime of $\Sigma_e$, being only sensitive to 
the branching ratio of the decay of $\Sigma_e$ itself into DM, which is kept fixed 
in the setup chosen. The two DM production processes occur at rather different 
time scales thus justifying the analytic computation of the DM relic density as 
the sum of the two indipendent contributions. 
In the second plot we adopted $m_{\Sigma_e}$ as varying parameter while the couplings 
$\lambda$ and $\lambda^{'}$ have been fixed to $10^{-12}$. The FIMP mechanism, 
in this case, dominates the DM relic density at lower values of $m_{\Sigma_e}$ because 
there the high value of its annihilation cross section suppresses the SuperWIMP 
contribution. The latter grows in importance by increasing $m_{\Sigma_e}$ and becomes 
dominant above the TeV scale. 
Since no significant deviations in the numerical computation have been found with respect 
to our analytic estimates (see again the appendix for clarifications) 
then we will keep referring to the latter over the rest of the paper.

We can now verify the impact of the relic density constraint on the model. 
As just shown, in the case of colored $\Sigma_f$ the SuperWIMP contribution is 
substantially irrelevant for masses within the LHC range. We can thus determine
 the coupling $\lambda$ from the FIMP contribution (\ref{eq:Freezein}) as:
\begin{equation}
\label{eq:lambda_FIMP}
\lambda \simeq 1.59\times 10^{-12} {x}^{-1/2}{\left(\frac{g_{*}}{100}\right)}^{3/4}
{\left(\frac{\Omega^{\rm FI}h^2}{0.11}\right)}^{1/2}g_\Sigma^{-1/2}   
\end{equation}
After fixing this coupling, we compute the DM lifetime in terms of the remaining 
parameters and show it in fig.~(\ref{fig:FIMP}). Notice that the limits considered 
for ID are valid only for DM masses above 1 GeV. Indeed, for lower values of the DM 
mass there are no bounds from anti-protons fluxes and also the limits from Fermi 
disappear below 100 MeV since such low values of the mass lie below its threshold. 
On the other  hand, weaker limits are still present from the gamma-rays continuum, e.g 
from EGRET \cite{Sreekumar:1997un}.

\begin{figure}[htb]
\subfloat{\includegraphics[width=6.5 cm, height= 6.5 cm, angle=360]{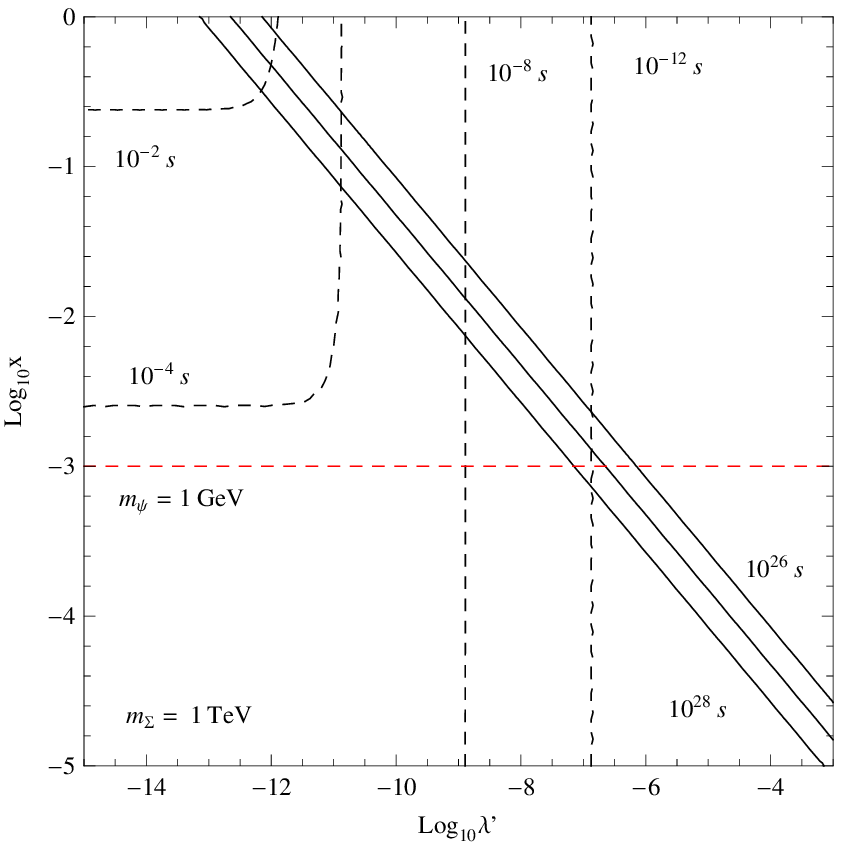}}
\hspace{10mm}
\subfloat{\includegraphics[width=6.5 cm, height= 6.5 cm, angle=360]{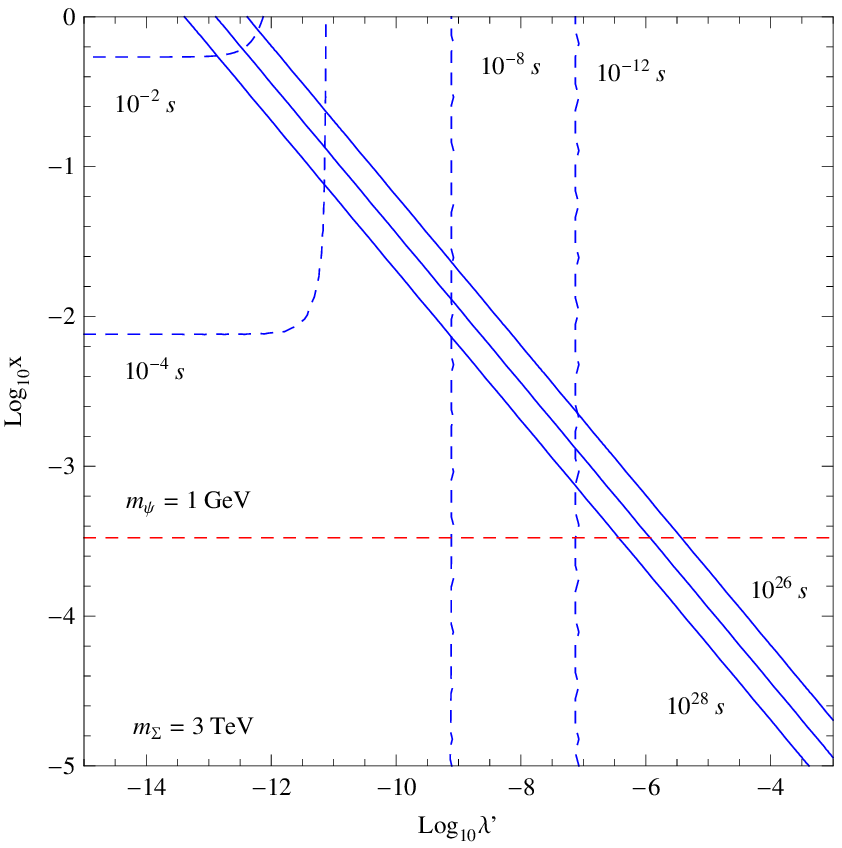}}
\caption{Solid lines: Contours of the DM lifetime as function of ${\mbox{Log}}_{10}x$ 
and ${\mbox{Log}}_{10}\lambda'$ for the values of $m_{\Sigma_f}$ reported in the plots. 
Dashed lines: contours of the $\Sigma_f$ lifetime. The dashed lines represent 
$m_\psi = 1 \mbox{GeV}$.}  
\label{fig:FIMP}
\end{figure}
The scalar field results now rather long-lived, hence we have to verify in addition 
that its decay does not affect Big-Bang Nucleosyntesis (BBN). 
In fig.~(\ref{fig:FIMP}) we have thus reported the values of the $\Sigma_f$ lifetimes 
varying the relevant quantities evidencing that the scalar decays before the onset 
of BBN over most of the parameter space.

In the case of leptonic models both FIMP and superWIMP mechanisms can play a relevant 
role, for masses of the scalar field within the LHC reach, in the determination of the 
DM relic density. This is displayed in fig.~(\ref{fig:xlr}), where we have given
the contours of the cosmological value of the DM relic density as function of 
$\lambda$ and $\lambda'$ for few sample values of $x$ and fixing 
$m_{\Sigma_{l,e}}=1 \mbox{TeV}$.  The dark matter relic density is dominated by the 
FIMP mechanism, therefore independent from $\lambda'$, at low x (where the SuperWIMP 
contribution is suppressed) and higher values of $\lambda$. The SuperWIMP contribution 
emerges at lower values of $\lambda$ for which, instead, the FIMP contribution is 
suppressed, but only as long as the ratio of masses $x$ and $BR(\Sigma_f \rightarrow \psi f)$ are not too small.

\begin{figure}[htb]
\begin{center}
\subfloat{\includegraphics[width=6.5 cm, height= 6.5 cm, angle=360]{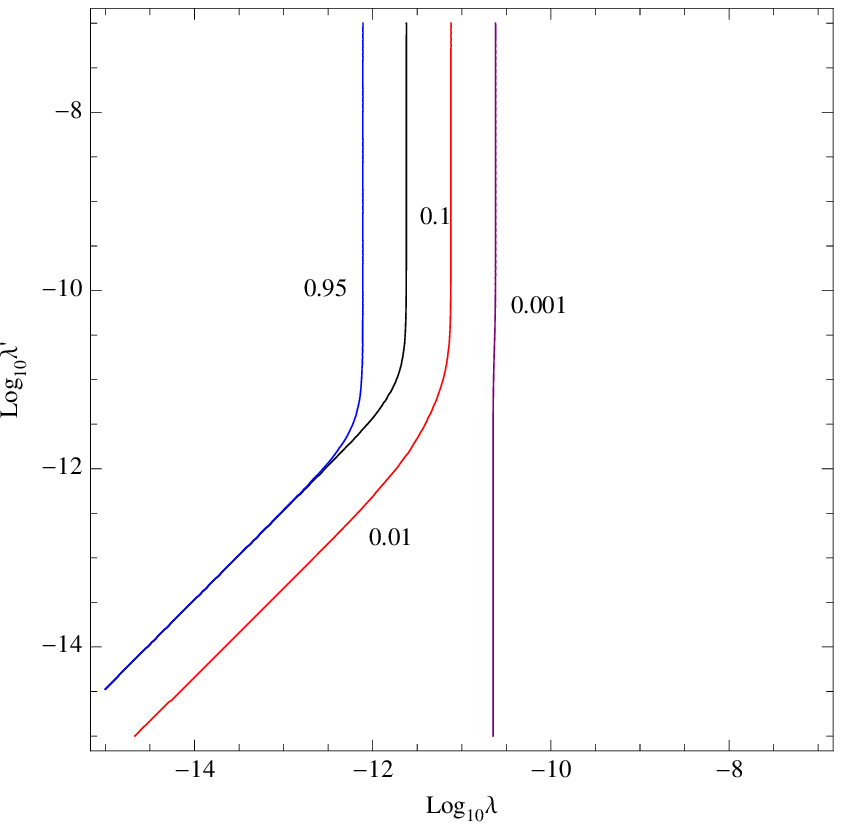}}
\hspace{10 mm}
\subfloat{\includegraphics[width=6.5 cm, height= 6.5cm, angle=360]{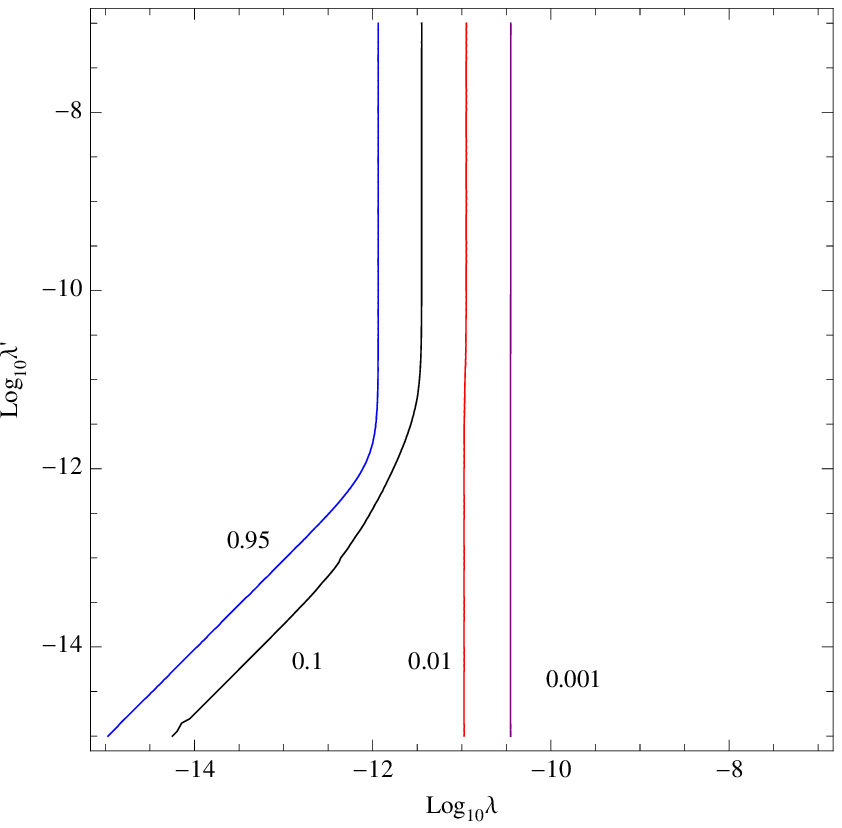}}
\end{center}
\caption{Contours of the thermal value of the DM relic density in the case of leptonic model with SU(2) doublet (left plot) and singlet (right plot) for the values of $x$ reported. The mass of the scalar have been fixed to 1 TeV.}
\label{fig:xlr}
\end{figure}

We have then compared the regions matching the correct DM relic density with
the regions giving a possible signal in indirect detection. For definiteness we 
have again fixed $m_{\Sigma_f}= 1$ TeV and considered two sample scenarios, 
corresponding to the values $x=0.1$ and $x=0.001$, for each of the two leptonic models. 
The upper panels of fig.~(\ref{fig:FISW}) report the results of our investigation 
for the case $x=0.1$, with the $\Sigma_l$ and $\Sigma_e$ models on the left and on the right
respectively. In the upper left panel we can see that our reference value of 
$10^{27}\mbox{s}$ of the DM lifetime is achieved when the correct value of the DM relic 
density is mainly determined by the SuperWIMP mechanism, while in the right panel
both FIMP and SuperWIMP mechanisms are relevant in that region.
In the plots we give as well the values of $BR(\Sigma_f \rightarrow \psi f)$ and
we see that the cosmological and indirect detection interesting regions  
around $\lambda' \sim 10^{-11}$ and $\lambda \sim 10^{-12}$ allow for
branching ratios of ten percent, so that possibly both $\Sigma_f$ decays may 
be within the reach of LHC.
Indeed since the FIMP mechanism  forbids values of $\lambda$ greater than order 
of $10^{-12}$, it implies for $\lambda^{'}$ a value not much different when 
a DM lifetime close to experimental sensitivity is assumed, as can be seen 
from fig.~(\ref{fig:Iplot}). 

In the second sample scenario considered, i.e. $x=0.001$ (see lower panels of 
fig.~(\ref{fig:FISW})) experimentally interesting values of the DM lifetimes 
are achieved where the FIMP mechanism accounts for its whole relic density. 
This is a consequence of the strong dependence of the DM lifetime on $m_\psi$. 
Indeed, by decreasing its value, the DM lifetime is strongly suppressed and
the only way to compensate this effect is by increasing $\lambda^{'}$ since 
$\lambda$, as given by eq.~(\ref{eq:lambda_FIMP}), can increase at most 
as $1/\sqrt{x}$. This however implies an extreme suppression of 
$BR(\Sigma_f \rightarrow \psi f)$ which makes irrelevant the SuperWIMP contribution.   

 \begin{figure}[htb]
\begin{center}
\subfloat{\includegraphics[width=6.5 cm, height= 6.5 cm, angle=360]{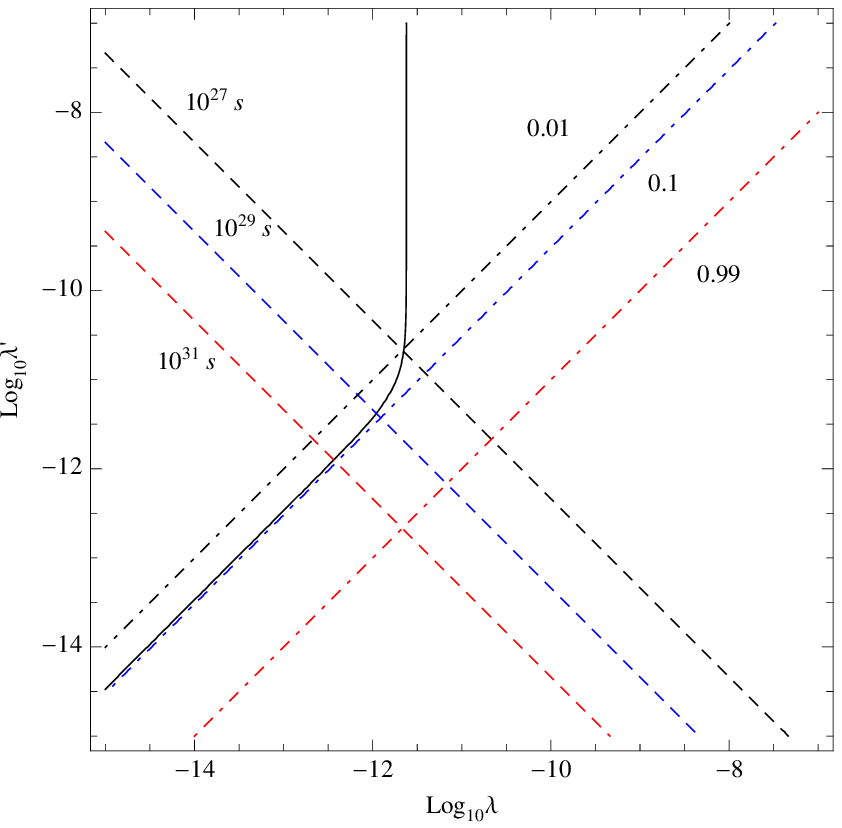}}
\hspace{2 mm}
\subfloat{\includegraphics[width= 6.5cm, height= 6.5cm, angle=360]{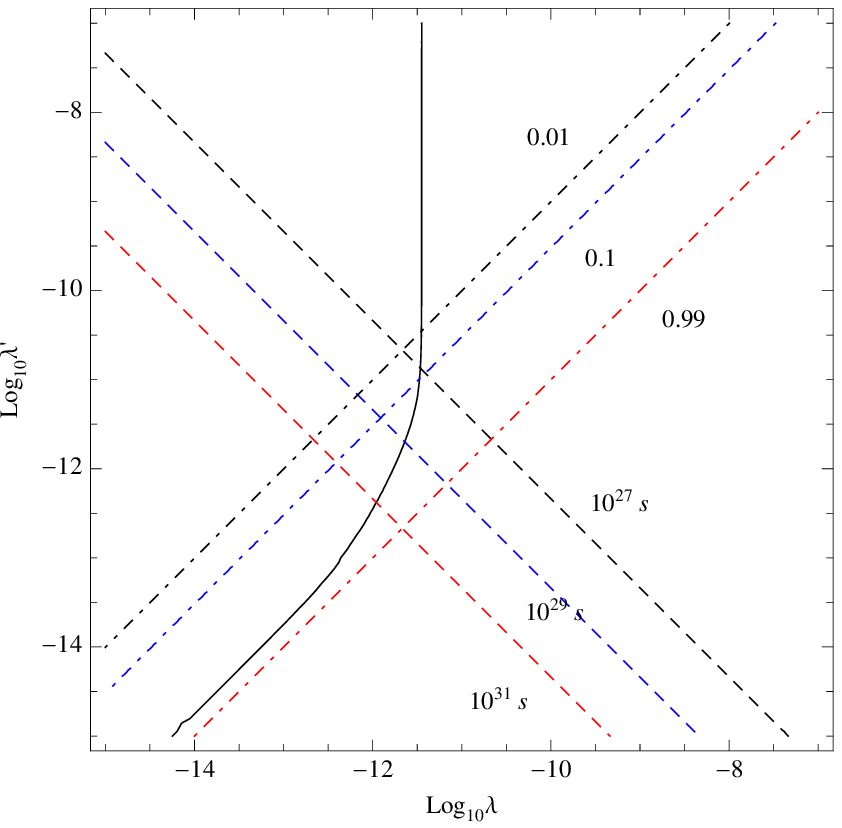}}\\
\subfloat{\includegraphics[width= 6.5cm, height= 6.5cm, angle=360]{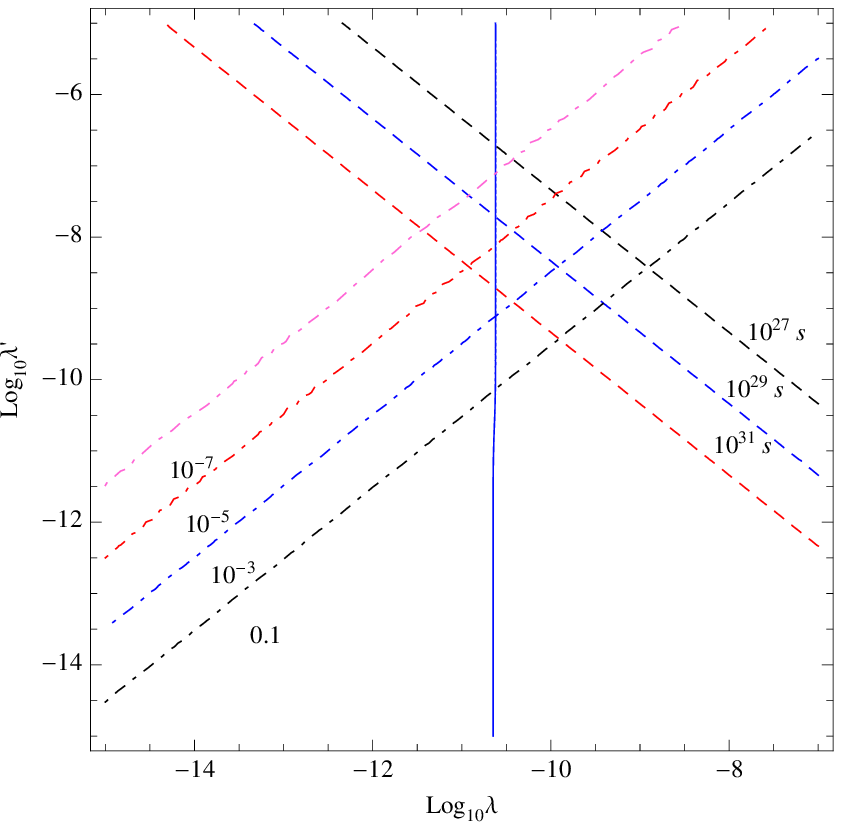}}
\hspace{2 mm}
\subfloat{\includegraphics[width= 6.5cm, height= 6.5cm, angle=360]{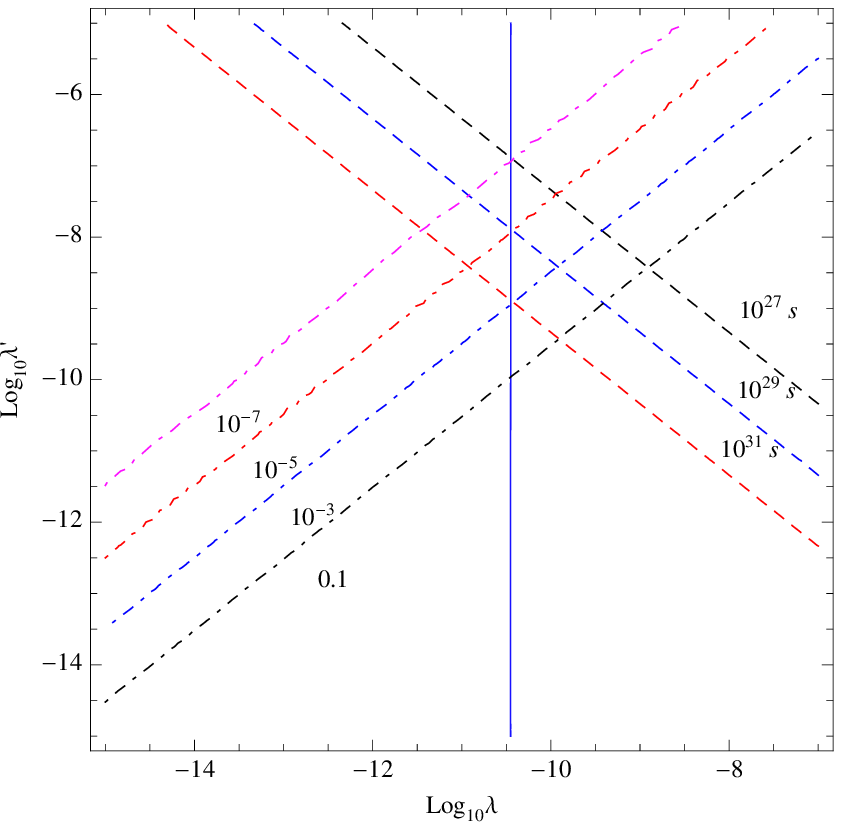}}
\end{center}
\caption{Solid lines: curves of the cosmologically favored value of the DM relic density. 
Dashed lines: curves of the DM lifetime associated to the values reported in the plot. 
Dash-dot lines: Branching ratio of decay of $\Sigma_f$ into DM and one fermion. The values 
are reported as well in the plot. Left column refers to the case of an SU(2) doublet. 
Upper plots are obtained for $x=0.1$ Lower plots for $x=0.001$. $m_{\Sigma_f}$ is fixed at 1 TeV.}
\label{fig:FISW}
\end{figure}

Although not strictly relevant for LHC phenomenology, we now discuss, as final mention, scenarios of heavy (namely above multi-TeV scale) dark matter and $\Sigma_f$ particles since they will be 
object of the next future CTA \cite{Cirelli:2012ut} and AMS-02 \cite{Cirelli:2013hv} searches. In these scenarios DM is generated 
through the SuperWIMP mechanism since the high values of $m_{\Sigma_f}$ do not suppress anymore 
its relic density even in the case of strong interactions. 
In fig.(\ref{fig:SWC}) we consider three sample values of the pair $(x,m_{\Sigma_f})$, 
i.e. $(0.5,3\,\mbox{TeV})$ $(0.1,10\,\mbox{TeV})$ and 
$(0.1, 100\,\mbox{TeV})$, for the scenario $\Sigma_d$. As evident, cosmologically viable candidates can have lifetimes within the experimental 
sensitivity and, at the same time, BBN is safe, as shown 
by the contours of the $\Sigma_d$ lifetimes. 
\begin{figure}[htb]
\begin{center}
\subfloat{\includegraphics[width= 6.5 cm, height= 6.5 cm, angle=360]{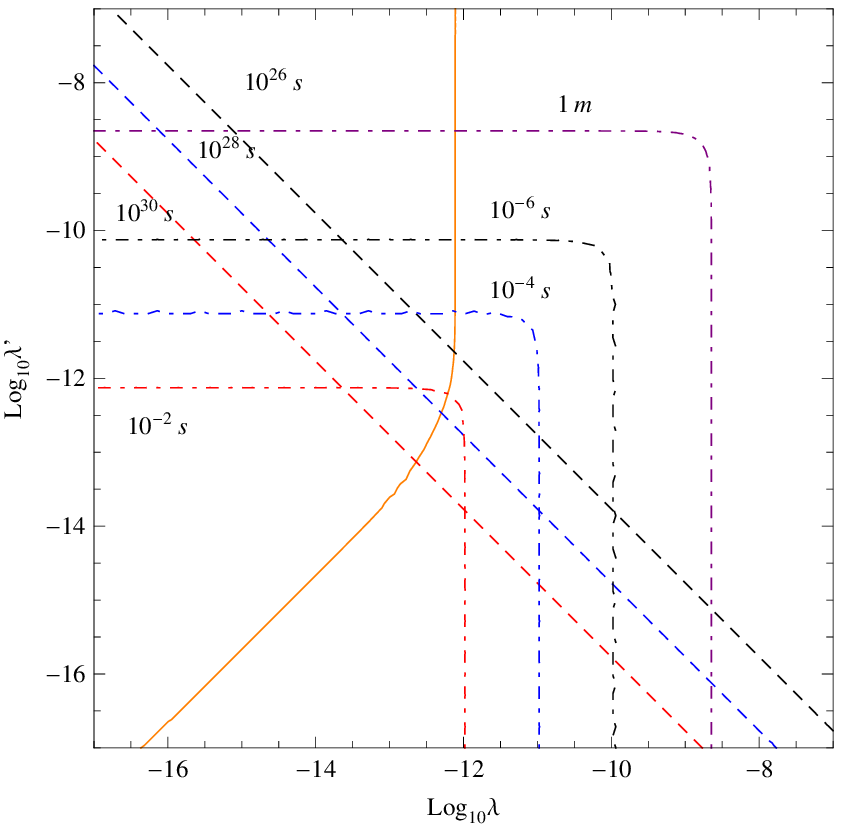}}
\hspace{10 mm}
\subfloat{\includegraphics[width= 6.5cm, height= 6.5cm, angle=360]{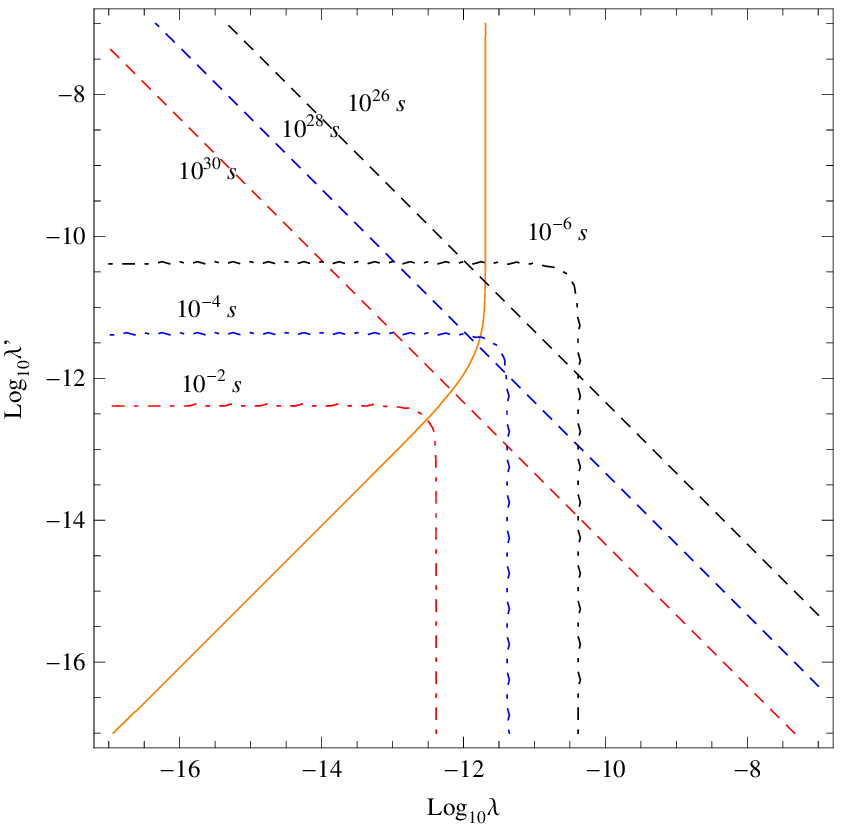}}\\
\subfloat{\includegraphics[width= 6.5 cm, height= 6.5 cm, angle=360]{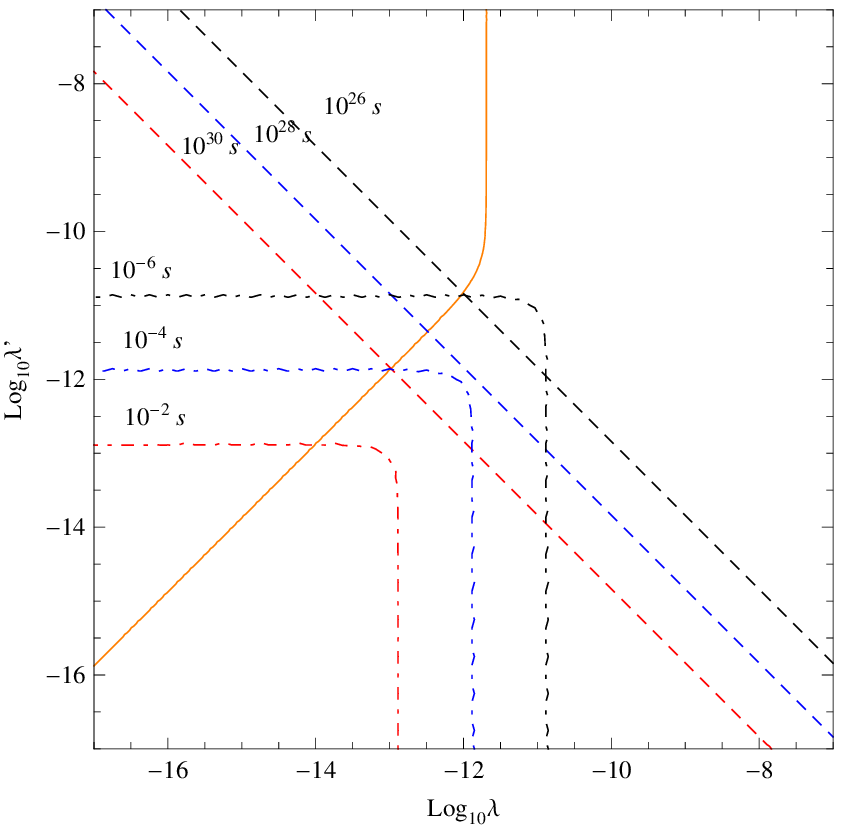}}
\end{center}
\caption{Solid lines: curves of the WMAP value of the DM relic density. Dashed lines: curves of the DM lifetime associated 
to the values reported in the plot. Dash-dot lines:Values of the $\Sigma_d$ lifetime reported in the plots. The three scenarios refer to the values of 
$(x,m_{\Sigma_d})$ of, respectively, $(0.5,3\,\mbox{TeV})$ $(0.1,10\,\mbox{TeV})$ and $(0.1, 100\,\mbox{TeV})$. In the first case the mass of $\Sigma_d$ 
is actually still within the LHC reach although its decay length (see magenta dash-dotted line) is too high for guaranteeing observation of its decays.}
\label{fig:SWC}
\end{figure}

\subsection{WIMP regime}

In this subsection we will briefly revisit the case when the Dark matter abundance 
is generated through the WIMP mechanism. Along our discussion we will stick to hadronic 
models since they are the most strongly constrained by the recent experimental results. 
Leptonic models are much more weakly constrained and will be very shortly mentioned 
in the next sections. 

According to the conventional WIMP paradigm the dark matter is in thermal equilibrium 
until it becomes non-relativistic and its relic density is set by the pair annihilation 
cross section. A Majorana fermion SM singlet as our DM candidate $\psi $ annihilates 
dominantly into fermion pairs. At low velocities the thermally averaged cross section 
can be expressed as:
\begin{equation}
\langle \sigma v \rangle \simeq \frac{9 \lambda^4 x^2}{64 \pi {\left(1+x^2\right)}^2 m^2_{\Sigma_f}} r^{-1}
\end{equation}
where we have defined $r=m_\psi/T$.
From this it is possible to compute the DM relic density according the well known expression \cite{Gondolo:1990dk}:

\begin{align}
&\Omega_\psi^{\rm WIMP} h^2 = 8.76 \times 10^{-11} {\mbox{GeV}}^{-2}{\left[  \int_{T_{0}}^{T_{\rm f}} \bar{g}_{\rm eff} \langle \sigma v \rangle \frac{dT}{m_\psi} \right]}^{-1} \nonumber\\
& \approx 3.9 \times 10^{-4} \frac{r_{\rm f.o.}^2}{x^2 \lambda^4}{\left(\frac{m_{\Sigma_f}}{1 \mbox{TeV}}\right)}^{2}{\left(\frac{g_{\rm eff}^{1/2}}{10}\right)}^{-1}
\end{align}
The observed value of the DM relic density $\Omega h^2 \simeq 0.11$ is achieved for: 
 \begin{equation}
 \lambda_{\rm WIMP} \approx 0.8\, x^{-1/2} {\left(\frac{m_{\Sigma_f}}{1 \mbox{TeV}}\right)}^{1/2}
 {\left(\frac{\Omega_\psi^{\rm WIMP}h^2}{0.11}\right)}^{-1/4}
 \label{eq:lambda_WIMP}
 \end{equation}
 Considering values of $m_{\Sigma_f}$, in agreement with 
collider limits, as reviewed in the last section, the correct relic can be achieved only 
for $\lambda$ greater than one. Lower values of $\lambda$ may be viable when $x$ gets 
close to 1 and also coannihilation effects induced by the the scalar field $\Sigma_f$ 
become important. As already stated, we will not include this case in our discussion 
since is has been already investigated in detail in~\cite{Garny:2012eb}. 
From the relation (\ref{eq:lambda_WIMP}) we can determine, as function of the DM mass, 
the maximal value of $m_{\Sigma_f}$ compatible with perturbative interactions, 
namely $\lambda \le 4 \pi$:
 \begin{equation}
  m_{\Sigma_f} \lesssim 5\,\mbox{TeV} {\left(\frac{m_\psi}{100\mbox{GeV}}\right)}^{1/2}
 \end{equation}
 
The high value of $\lambda$ required to obtain the right relic density 
may be responsible as well of a sizable signal at direct detection experiments. 
The strongest constraints come from spin-independent (SI) interactions which 
can be described by an effective lagrangian analogous to the one defined 
e.g. in \cite{Drees:1993bu} (see also \cite{Garny:2012eb} for a recent computation
 of the relevant cross-section).  

 The lagrangian (\ref{eq:DMint}) is responsible as well of spin dependent interactions, 
however, as shown in \cite{Fox:2011pm}, 
the limits obtained are much weaker than the ones coming from collider searches and therefore will be not discussed further.

Having obtained a determination, as function of $x$ and $m_{\Sigma_f}$, of the coupling $\lambda$ we can infer the range of values of the coupling 
$\lambda^{'}$ accessible in ID experiments from the expression (\ref{eq:Gamma_DM}) of the DM decay rate. Assuming the limit value of $10^{27}\mbox{s}$, 
for $m_{\psi} \gtrsim 1\mbox{GeV}$, for the DM lifetime,  the coupling $\lambda^{'}$ is required to be of the order of $10^{-22}$, 
and then results irrelevant for the other aspects for the theory.

We finally mention that, in addition to the bounds from DM decay already discussed, one should also consider the ones coming for the detection 
of the DM annihilation processes. 
In this case the most stringent limits come from the detection of $\gamma$-rays which, especially in the case of annihilation into quarks, 
are already able to severely constraint or even rule-out light, i.e. order of tens of GeV, 
thermal relics (see for example \cite{fortheFermiLAT:2013naa}).

In the next sections we will investigate which kind of collider phenomenology can be observed 
for the values of $\lambda$ and $\lambda^{'}$ selected by relic density and ID constraints. 
Before doing this we will briefly comment on other general constraints which may apply to 
the scenarios under investigation.

\section{Other constraints}

The operators relating the field $\Sigma_f$ and the SM fields do not preserve lepton and/or baryon number and therefore suffer additional constraints apart to cosmology and DM detection. 
Indeed a rather large variety of flavor violating processes may be induced, depending 
on which kind of coupling between the scalar field and the SM is switched on. 
Given the analogy with RPV SUSY realizations, one can substantially adopt the experimental limits 
existing for this class of models. A rather exhaustive list can be found in \cite{Barbier:2004ez}. 
The impact of such limits is rather model dependent relying, in particular, on an eventual 
flavor structure of the coupling among the scalar field and the SM fields. In most cases, 
however, these limits are much weaker than the ones imposed by ID and thus result irrelevant 
to our discussion.

The only exception is for the case in which $\Sigma_f$ behaves like a $\tilde{d}_R$ since 
the combination $\lambda^{'} \lambda^{''}$ induces proton decay. Assuming that 
$\Sigma_d$ couples democratically with all quark flavors, we obtain a limit even as 
stringent as $| \lambda'\,\lambda''| < 10^{-27}$ which can be instead reduced down 
to $O(10^{-9})$ \cite{Smirnov:1996bg} if a flavor structure is assumed for these couplings.
These limits can be thus relevant in FIMP scenarios but their actual impact is again 
model dependent. On purely phenomenological grounds these can be easily overcome by 
assuming a suitable hierarchy among the couplings $\lambda'$ and $\lambda''$. 

\section{Prospects for LHC detection}

In this section we will examine whether the information provided by cosmology can be translated 
into predictions of collider detection. The first part of our discussion will be devoted 
to the general features shared by the various realizations, namely hadronic and leptonic,  
defined above, while at a second time we will specialize on the specific setups 
indicating possible peculiar signatures.

On general grounds we expect the most important signatures to originate from the pair 
production and subsequent decay of $\Sigma_f$ particles. 
In the previous sections we have shown that, depending on the cosmology, different ranges 
of the couplings $\lambda$ and $\lambda'$ are possible and, consequently, the $\Sigma_f$ 
field features a broad range of lifetimes, leading to prompt as well as displaced decays 
or, alternatively, metastable tracks.
Interestingly, the ranges of couplings $\lambda$ and $\lambda'$ set by the requirements 
of a viable DM candidate, give different collider phenomenologies and thus can be 
discriminated by an eventual signal. 

The production rate of the field $\Sigma_f$ is determined by gauge interactions and 
is essentially fixed by its mass. As a consequence of the direct coupling of $\Sigma_f$ 
with quarks and gluons, higher production rates and, in turn, stronger constraints 
are obtained in the case of hadronic models compared to leptonic ones. 

On general grounds we distinguish the following scenarios:
\begin{itemize}
\item The field $\Sigma_f$ decays promptly inside the detector if its decay length $c \tau_\Sigma \gamma$ is much lower than $O(100 \mu\mbox{m})$.
\item For  $ O(100 \mu\mbox{m}) \lesssim c \tau_\Sigma \gamma \lesssim O(10\,m)$ $\Sigma_f$ 
decays still inside the detector but the decay vertex is displaced with respect to the primary one (this kind of scenario is
discussed, for example, in \cite{Chang:2009sv,Graham:2012th}).
\item For longer lifetimes there is no in-flight decay inside the detectors. 
The field $\Sigma_f$ can be nonetheless observed as massive charged track 
(see e.g.\cite{ATLAS:2012jp} for an example of search). In addition has been proposed the possibility of detecting 
long lived particles stopped in the detector by looking at their decay at rest during the periods in which no $pp$ collision take 
place \cite{Asai:2009ka}.  
\end{itemize}

This is actually a too rigid classification since it does not take into account possible 
uncertainties and detector effects which affect the possible detection of the decay products. 
In reality, the values of the decay length discriminating the different cases 
depend also on the type of decay products and thus result model dependent~\cite{Graham:2012th}. 
We remark that we will pursue a qualitative analysis relying on the computation of the 
$\Sigma_f$ lifetimes without taking into account kinematical quantities (like boost factors) 
affecting the actual detection of decays of the scalar field. These quantitative aspects 
will be object of a dedicated study.

In case of possible detection of the $\Sigma_f$ decay products we identify two event categories, 
depending on which of the two kinds of couplings, namely $\lambda$ and $\lambda'$, dominates. 
In the first case the field $\Sigma_f$ decays into a quark or a lepton in addition to the 
DM state $\psi$ implying a sizable amount of missing energy. In the opposite case instead 
we have decays into only standard model states with a drastic reduction of the amount of 
missing energy. The specific signatures are more model dependent and will be discussed 
in more detail in the next subsection.

The WIMP scenario is characterized by a coupling $\lambda$ of order 1 or even larger 
compared to a substantially irrelevant (apart from DM decays) $\lambda^{'}$.  As a 
consequence pair produced $\Sigma_f$ feature prompt decays into a quark or a lepton, 
depending on its quantum numbers, and the DM giving the classical WIMP signature
of missing energy in all events. 

More involved is instead the other scenario. In the case of the colored $\Sigma_{f=q,u,d}$
within LHC reach, we have dominance of the FIMP mechanism. By reverting eq.~(\ref{eq:lambda_FIMP}) 
we can express its decay length in terms of the DM relic density giving: 
  \begin{equation}
 l_{\Sigma,DM}= 2.1 \times 10^5 \mbox{m}\, g_\Sigma x\; 
{\left(\frac{m_{\Sigma_{f}}}{1 \mbox{TeV}}\right)}^{-1} 
{\left(\frac{\Omega_{CDM} h^2}{0.11}\right)}^{-1}
{\left(\frac{g_{*}}{100}\right)}^{-3/2}
 \end{equation}  
It is therefore evident that the decay length into DM largely exceeds the size of the 
detector except for very low values of $x$, namely less than $10^{-4}$. 
On the other hand, by imposing a ID signal (see fig.~\ref{fig:FIMP}) in the near future, 
we obtain an enhancement of the the coupling $\lambda^{'}$, responsible of decays of 
$\Sigma_f$ into SM states only, of several orders of magnitude with respect to $\lambda$. 

In order to have an estimate of the decay length associated to the decay mediated by 
$\lambda^{'}$, we combine the expression of the DM lifetime, eq.~(\ref{eq:Gamma_DM}), 
with eq.~(\ref{eq:lambda_FIMP}) and get $\lambda'$ as function of $x$ and $m_{\Sigma_f}$: 
\begin{equation}
\label{eq:lambda_prime}
\lambda^{\prime} \simeq 0.91 \times 10^{-12}\; x^{-2}
{\left(\frac{g_{*}}{100}\right)}^{-3/4}{\left(\frac{m_{\Sigma_{f}}}{1\mbox{TeV}}\right)}^{-1/2}
g_\Sigma^{1/2}{\left(\frac{\tau_\psi}{10^{27}\mbox{s}}\right)}^{-1/2}
{\left(\frac{\Omega_{\rm CDM}h^2}{0.11}\right)}^{-1/2}\; ,
\end{equation}
corresponding to a decay length:
\begin{equation}
l_{\Sigma, SM} \simeq 55 \,\mbox{m}\, \frac{1}{g_\Sigma} 
{\left(\frac{m_{\Sigma_f}}{1\mbox{TeV}}\right)}^{-4} 
{\left(\frac{m_\psi}{10\mbox{GeV}}\right)}^{4} 
\left(\frac{\tau_\psi}{{10}^{27}\mbox{s}}\right)
\left(\frac{\Omega_{CDM} h^2}{0.11}\right) 
{\left(\frac{g_{*}}{100}\right)}^{3/2}\; 
\end{equation}
mostly within the detector for DM much lighter than the scalar field. 
Since we are interested in identifying scenarios of possible contemporary detection of the two decay channels it 
is instructive to compare the previous expressions of the decay lenghts from which we can infer the relative branching ratios 
of the two channels:
\begin{equation}
\label{eq:ratio_lenghts}
\frac{l_{\Sigma, DM}}{l_{\Sigma, SM}} \equiv \frac{BR(\Sigma_f \rightarrow SM)}{BR(\Sigma_f \rightarrow \psi f)}
\simeq 38\, g_\Sigma^2 {\left(\frac{x}{0.01}\right)}^{-3}
{\left(\frac{m_{\Sigma_f}}{1\mbox{TeV}}\right)}^{-1}
{\left(\frac{\tau_\psi}{{10}^{27}\mbox{s}}\right)}^{-1}
{\left(\frac{\Omega_{CDM} h^2}{0.11}\right)}^{-2}
{\left(\frac{g_{*}}{100}\right)}^{-3}
\end{equation}

 \begin{figure}[htb]
 \begin{center}
\subfloat{\includegraphics[width=6.5 cm, height= 6 cm, angle=360]{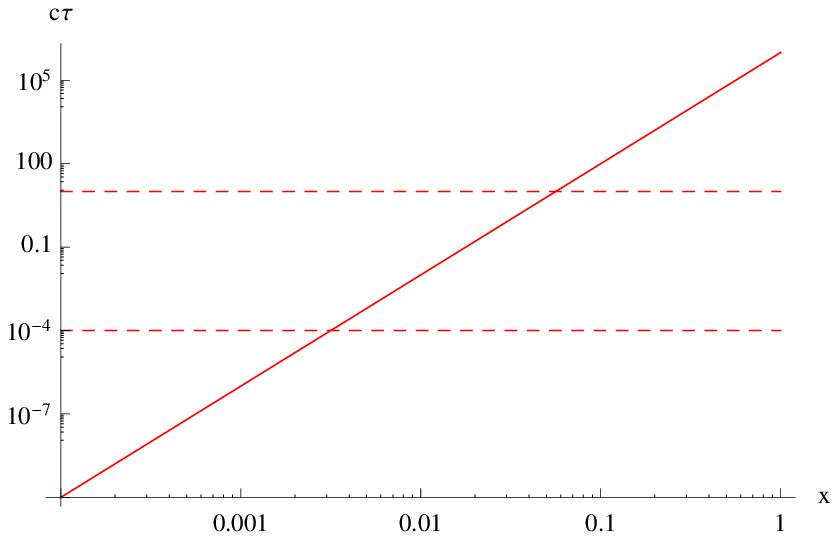}}
\hspace{2mm}
\subfloat{\includegraphics[width=6 cm, height= 6 cm, angle=360]{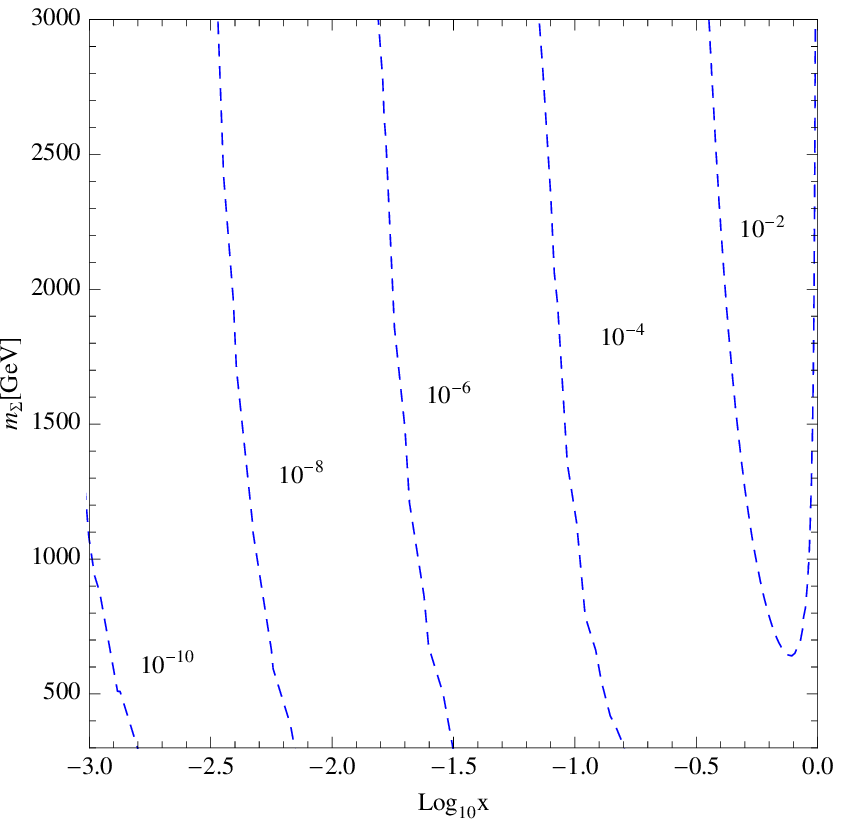}}
\end{center}
\caption{Decay length of the field $\Sigma_f$ into only SM states (left panel) and 
branching ratio of decays into DM (right panel). 
The couplings have been fixed according eq.~(\ref{eq:lambda_FIMP}) and (\ref{eq:lambda_prime}).}  
\label{fig:ctau}
\end{figure}

In the left panel of fig.~(\ref{fig:ctau}) we have reported the value of the decay length 
of the field $\Sigma_f$ into only SM particles as a function of $x$. Considering a reference 
value of $m_{\Sigma_{q,u,d}}$ of the order of 1 TeV, we expect to observe displaced vertices
for dark matter masses down to around 1 GeV while for lower values the decay can be prompt. 
In the right panel of fig.~(\ref{fig:ctau}) we report as well the values of 
$BR(\Sigma_{q,u,d} \rightarrow q \psi)$ in the plane $(x,m_{\Sigma_f})$. This branching ratio reaches 
its maximal value for nearly degenerate masses where it can exceed the order of percent. However
 in that region the decay length of $\Sigma_f$ is very large. For example for $x=0.5$ and 
 $m_{\Sigma_d}=2.5\,\mbox{TeV}$ it is around $6 \times 10^5\mbox{m}$.

An analogous outcome is obtained for the leptonic models. 
In fig.~(\ref{fig:ctauEW}) we have plotted the total decay length of the scalar $\Sigma_l$
SU(2) doublet together with some relevant values of the DM lifetime. For DM masses down 
to 1 GeV the decays of $\Sigma_f$ are expected to be observed at most as displaced vertices. 
Comparing this figure with (\ref{fig:FISW}) we notice again that values of the branching 
ratio of decay of the scalar field into DM above ten percent are disfavored. 
 
 \begin{figure}[htb]
 \begin{center}
 \subfloat{\includegraphics[width= 6 cm, height= 6 cm, angle=360]{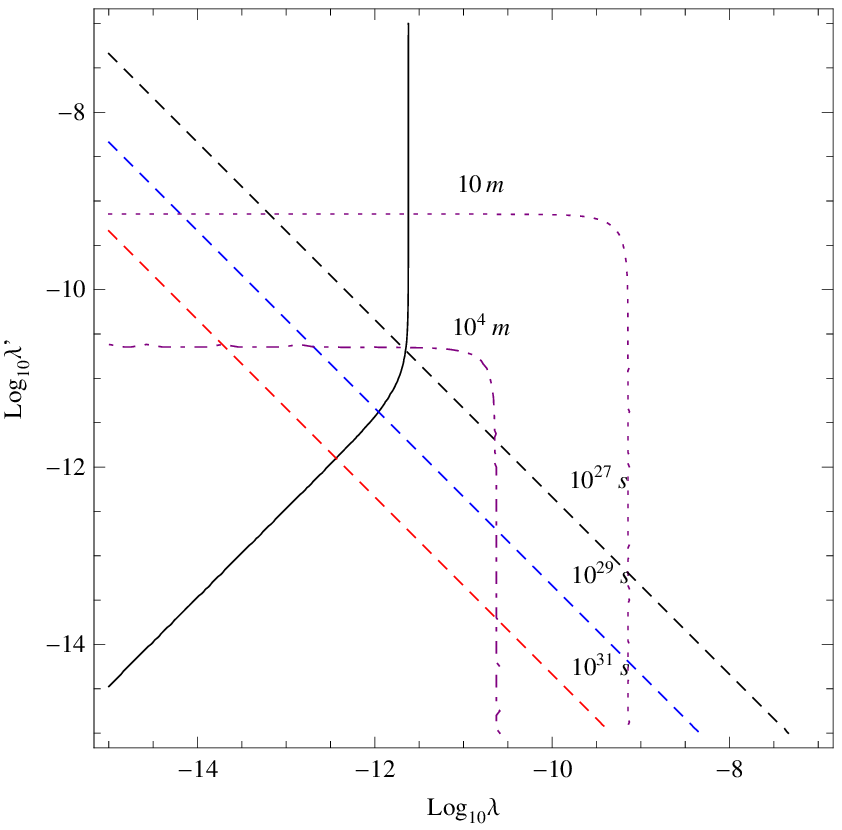}}
 \hspace{2mm}
 \subfloat{\includegraphics[width= 6 cm, height= 6 cm, angle=360]{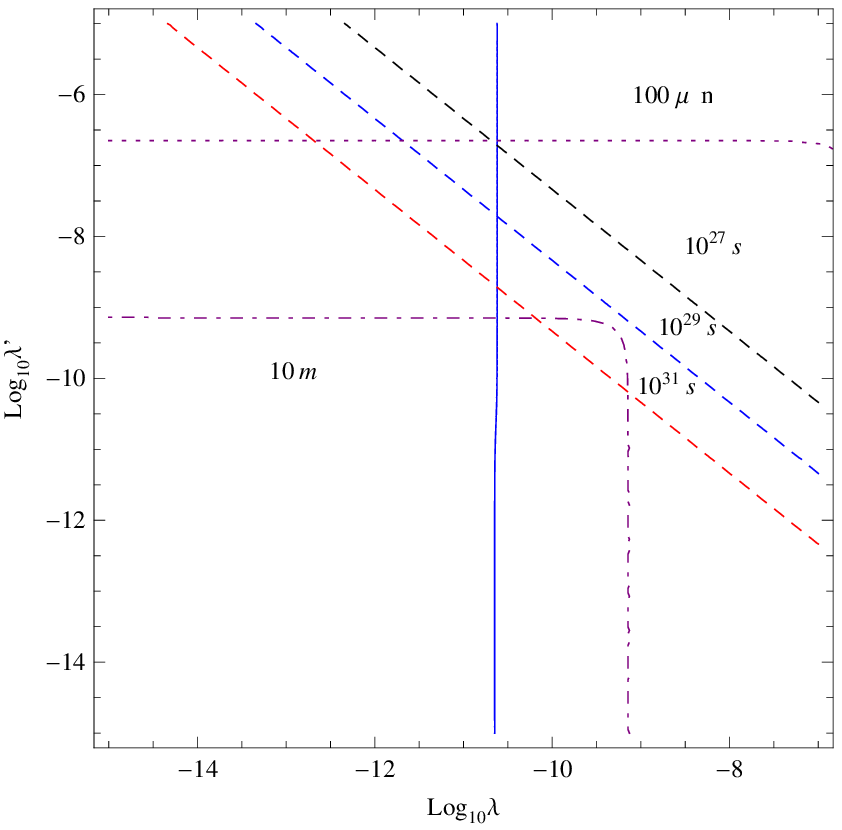}}
 \end{center}
 \caption{Solid lines: cosmological value of the DM relic density. Dashed lines: values of the DM lifetime as reported in the plot. 
 Left and right panel refers respectively to $x=0.1$ and $x=0.001$. $m_{\Sigma_f}$ have been fixed to 1 TeV. 
 For definiteness we have assumed $\Sigma_f$ a SU(2) doublet. In the plots are reported as well, through magenta lines, the reference values of
 $100\,\mu\mbox{m}$ and $10\mbox{m}$. In the regions above the former value the scalar field $\Sigma_f$ can feature prompt decays while below the 
 $10\mbox{m}$ curve it is detector stable. In the intermediate regions, instead, it is possible to observe displaced vertices. In the left plot the viable
 values of the DM lifetime all lie in the region in which $\Sigma_f$ is detector stable. In particular a DM candidate with correct relic density and 
 lifetime of around $10^{27}\mbox{s}$ is associated to a value of the order of $10^4\,\mbox{m}$ of the $\Sigma_f$'s decay lenght.}
 \label{fig:ctauEW}
 \end{figure}
 
 It is anyway possible to have comparable values of the rates of decay of the scalar 
field into DM and only SM states, together with decay length compatible with a detection 
at the LHC, for very light dark matter candidates. Fig.~(\ref{fig:noid}) shows the example
of a 1 TeV SU(2) doublet and SU(3) singlet scalar and a DM mass of 1 MeV (thus corresponding to $x=10^{-6}$). As evident, 
the two branching ratio of decays have comparable values and the relative 
decay lengths fall in a range of values giving displaced vertices. On the other hand we see from eq.~(\ref{eq:ratio_lenghts}) that, 
for such low values of the DM mass (or $x$ equivalently), the two kind of decay lenghts can be comparable only for 
very long DM lifetimes\footnote{Notice that one has actually to impose an even more severe condition since, for such low masses, 
DM decay is induced at one-loop while \ref{eq:ratio_lenghts} is obtained for tree-level decays.}. 
In the example considered the dominant dark matter decay process is into $\gamma\,\nu$ 
which can be possibly manifest as a $\gamma$-ray line.
The combined searches of experiments like, CHANDRA and XMM, fix the current experimental sensitivity to around 
$10^{27 \div 28}$ s~\cite{Herder:2009im}. As shown by fig.~(\ref{fig:noid}), these values 
are hardly achievable in our particle physics setup unless the coupling $\lambda^{'}$ gets 
close to order one. This can originate tensions with constraints from rare flavor violating 
processes, although the latter are rather model dependent, and moreover forbids again any 
possibility of collider detection of processes related to the coupling $\lambda$.
 
 \begin{figure}[htb]
 \begin{center}
 \includegraphics[width= 6 cm, height = 6 cm, angle=360]{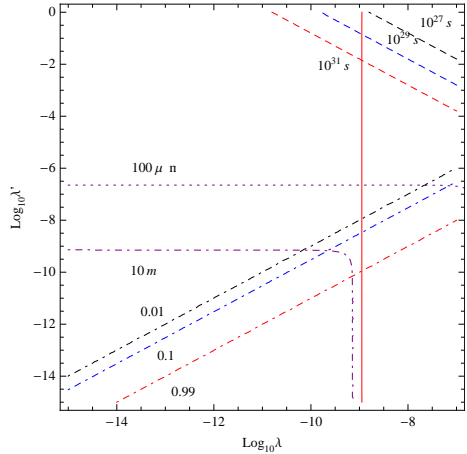}
 \end{center}
 \caption{Solid lines: cosmological value of the DM relic density. Dashed lines: values of the DM lifetime as reported in the plot. Dotted lines: values of 
 the $BR(\Sigma_f \rightarrow f\,\psi)$ as reported in the plot. Dot-Dashed lines values of $\Sigma_f$ decay length. $m_{\Sigma_f}$ have been fixed to 1 TeV 
 and $x=10^{-6}$. We have assumed $\Sigma_f$ a SU(2) doublet and a SU(3) singlet.}
\label{fig:noid}
\end{figure}

\subsection{LHC signatures}

In this subsection we will discuss in more detail the different realizations 
of our particle physics framework identifying the most relevant LHC signatures in each case. 

\subsubsection{FIMP/SuperWIMP scenario}

The FIMP/SuperWIMP scenario features the richest and yet partially unexplored phenomenology, 
given the different possible couplings as well as the range of lifetimes allowed by 
DM requirements. As discussed in the previous section, the collider prospects mostly rely 
on the couplings of the scalar field with SM fields only, apart for nearly degenerated
and large DM/$\Sigma_f$ masses, just below the LHC reach. 
As a consequence, events containing $\Sigma_f $ decays are characterized by only limited 
amount of missing energy, evading the most popular search strategies which are 
optimized for particle physics frameworks like R-parity conservative supersymmetry 
and rely on a stable massive particle, escaping detection, at the end of the decay chains.

In case of prompt decays, several alternative search strategies have been proposed,
mostly with the requirement of the presence of multiple leptons in the final state. 
These could be applied also in our models, in particular of type $\Sigma_{d,q,l,e}$.
On the contrary, models featuring only couplings with quarks, like $\Sigma_u$ in our 
framework, are, at the moment, hardly accessible at the LHC since their signals are, 
in absence of large amounts of missing energy, hidden by QCD backgrounds.

Among the searches currently carried out by the LHC collaborations, the ones which suit 
better our model are the searches for leptoquarks, i.e. exotic states carrying both lepton 
and baryon number and able to decay into both two types of channel. Searches of these 
states are performed in the channel $e\nu_e jj / \mu \nu_\mu jj$ 
\cite{cms:2012dnb} and $b \tau$\cite{Chatrchyan:2012sv} obtaining a lower bound on the 
mass which ranges from around 600 to 800 GeV, depending on the branching ratio of decay, 
in the first case and of around 500 GeV in the second. 
This kind of searches can be applied to the versions of our model featuring the operator 
$\lambda' lq\Sigma_f$, i.e. the $\Sigma_q$, $\Sigma_d$ and $\Sigma_l$ realizations. 
Another promising tool is the recent analysis by CMS of anomalous production of three 
leptons \cite{Aad:2012xsa} which can be, in principle, applied to the leptonic models.
 
Despite the analogy in the operatorial structure, current searches of R-parity violating 
supersymmetry are instead not straightforwardly applicable to our model since they are 
built to tag mainly RPV decays of neutralinos originating from RPC decays of pair produced 
superpartners (see e.g. \cite{Chatrchyan:2012mea, CMS-PAS-SUS-13-003})\footnote{In \cite{CMS-PAS-SUS-13-003} it is actually considered the case of 
RPV decays of a stop, 
however a four body decay is assumed.} and, consequently, require a greater number 
of objects in the final state than the one which can be provided by our simple model. 

A much clearer signature would be provided anyway by displaced decays away from the primary interaction 
vertices. Indeed, in this case, an eventual signal would suffer very suppressed or even 
irrelevant contribution from background processes although the reconstruction of this kind 
of events is often challenging~\cite{Graham:2012th}.  
This kind of signature is not fully explored since only few searches have been performed so far. 
The most recent is the one made by ATLAS in the context of RPV SUSY 
looking at displaced vertices with an additional muon \cite{Aad:2012zx} 
which has however the problem, already discussed, of requiring too many objects in the 
final state than the ones that can be accounted for in our model.   

We conclude with a final mention to the searches of detector stable particles which can instead
be relevant in FIMP/SuperWIMP scenarios at high values of the DM mass. Searches of long-lived 
particles have been presented both by ATLAS \cite{Atlas:2012vd} and, more recently, by 
CMS~\cite{Chatrchyan:2013oca} collaborations. Given the similarities we will adopt analogous 
limits to the ones derived for supersymmetric scalar particles, namely of the order of 
300-400 GeV for $\Sigma_{f=e,l}$ and slightly above 1 TeV for $\Sigma_{f=q,u,d}$.

\subsubsection{WIMP scenario}

In the WIMP scenario, the phenomenology is set only by the coupling $\lambda$, 
hence we distinguish two cases, namely the ones in which $\Sigma_f$ and $\psi$ are coupled 
to a quark or a lepton. The first scenario can be straightforwardly probed by mean of 
the multi-jet searches employed in the case or R-parity conserving supersymmetric models. 
Assuming flavor blind couplings, current searches exclude masses of the scalar slightly of 
800 GeV \cite{ATLAS-CONF-2012-109} assuming sizable mass splitting with the DM particle. 

However this is not the most stringent limit provided by collider physics. Indeed, 
the order 1 or more value of $\lambda$, needed to achieve the correct relic abundance, 
guarantees a sizable pair production rate of DM particles. The most powerful probe 
of direct pair production of DM are the search of events featuring the initial state 
radiation of jet or a photon. 

In the case of hadronic models the most updated limit comes from searches of monojet 
events performed by both the ATLAS and CMS collaborations whose results have been 
interpreted into bounds on several four-field effective 
operators~\cite{ATLAS:2012ky,ATLAS-CONF-2012-147,CMS-PAS-EXO-12-048}.  
Assuming the scalar field to be sufficiently heavier than the DM particle 
\footnote{Excluding eventual resonances the limits can be adopted as well when the assumption of contact interaction is not strictly valid anymore. 
In this case anyway they result conservative \cite{Fox:2011pm, Fox:2011fx}.}, 
several operators are allowed in our particle physics framework. 
First of all it is possible to obtain from the lagrangian (\ref{eq:DMint}) 
a four fermion interaction which can be translated, through a Fierz transformation,
into the following combination of effective operators:
\begin{equation} 
\frac{\lambda^2}{m_{\Sigma_f}^2} \bar{\psi}\gamma^{\mu}\gamma^{5}\psi \bar{q} \gamma_{\mu} \left(1 \pm \gamma^5\right) q 
\end{equation} 
In addition the interactions responsible for the DM scattering cross section with 
nucleons can be expressed in terms of an effective operator relating two DM particles 
with two gluons which result proportional to 
$\frac{\alpha_s \lambda^2}{4 m_{\Sigma_f}^3} \bar{\psi}\psi {\left(G^{\mu\nu}\right)}^2$. 
All these three operators in principle account for direct DM pair production. 
Assuming no cancellation effects we can adopt the most recent limits on the single 
operators as reported for example in \cite{CMS-PAS-EXO-12-048}. 
In particular, from these we obtain that the contribution of the gluon operator 
is irrelevant since it is suppressed by $\alpha_s$ and the higher power of 
$m_{\Sigma_f}$ with respect to the previous ones. Then we can translate the bounds
on those operators to our particle physics setup, and, for values of 
$\lambda$ of order one or greater,  we obtain a limit on $m_{\Sigma_f}$ as stringent 
as several TeV (see also fig.~(\ref{fig:summary_WIMP})). 
As will be seen in the next section, combining this limit with DD and the relic density 
requirement allows to exclude a WIMP DM candidate in large portions of the parameter space.

This statement is, however, strictly valid only assuming a coupling of $\Sigma_f$ 
with the light quark flavors.  Considering for example hierarchical couplings $\lambda$ 
in favor of third generation quarks, the limits form XENON and dark matter direct production 
are drastically reduced because of the low bottom and top content of the proton. 
In this scenario the strongest limits come from $\Sigma_{q,u,d}$ pair production.  
More precisely, masses of $\Sigma_f$ of  550-620 GeV are excluded in the case of decay 
into $b\psi$ \cite{ATLAS-CONF-2012-165} while masses between 300 and 700 GeV are excluded, 
for $\psi$ massless up to a mass of 150 GeV, in the case of decay into 
$t\psi$ \cite{ATLAS-CONF-2013-024}.

DM pair production has been also studied, in the case of leptonic models, through 
mono-photon searches at the LEP collider \cite{Abdallah:2003np}. In this case an upper limit on $m_{\Sigma_f}$ of around 
300 GeV have been derived \cite{Fox:2011fx} for DM masses up to 100 GeV. 
Comparable limits, namely of around 250-270 GeV from searches of $\Sigma_f$ pair 
production employing the results of LHC searches of events featuring two leptons 
and missing energy~\cite{CMS-PAS-SUS-12-022}.

\subsection{Discussion of results}

Before stating our conclusions in this section we are going to summarize our results 
highlighting the scenarios in which (possibly peculiar) collider and indirect detection 
signatures might be both present and their correlation can provide a guideline for 
future detection strategies.

On general grounds ID signals give information on the DM mass and on the global size 
of the combination $\lambda \lambda'/m_{\Sigma_f}^2$, while from an hypothetical LHC signal, 
it could be possible to discriminate the relative size of the two couplings, the mass of 
the scalar field and its quantum numbers. 
The clearest signature would be in any case a double collider detection of both the 
two $\Sigma_f$ decays into DM and a SM state and into SM particles only.  
In this favorable case, it would be possible to measure all the model parameters from 
LHC data and predict the DM indirect detection signal expected. At the same time it may 
be possible to test the FIMP/SuperWIMP production mechanism, similarly as it was proposed 
for the WIMP case~\cite{Baltz:2006fm,Bertone:2010rv,Bertone:2011pq}.
Unfortunately, such a double signal is not easily achieved in our scenario, together with an 
acceptable cosmological DM density and a DM lifetime 
within reach of ID observations. As shown in fig.~(\ref{fig:ctau}), in hadronic models 
the decay lengths of $\Sigma_f$ into DM and only SM particles can be close in value, compatibly within an 
observable DM lifetime, in a region at nearly degenerate DM and scalar masses
at the extreme energy reach of the collider. In this region however the $\Sigma_f$ decay 
length largely exceeds the dimension of the detector. In leptonic models comparable decay 
lenghts can be achieved for lower masses of the scalar field $\Sigma_f$ and not degenerate 
values of the DM mass (see .e.g. fig~\ref{fig:ctauEW}), but again their value exceeds the 
dimension of the detector.
Although disfavored, the double detection of the decay of $\Sigma_f$ in these cases 
cannot be completely ruled out. Indeed, as shown in \cite{Ishiwata:2008tp}, once the
production rate and the momentum distribution is properly taken into account, a residual 
population of particles decaying inside the detector may exist even for long lifetimes. In 
addition, as discussed e.g. in \cite{Asai:2009ka}, it is possible to detect even very late decays 
of charged particles in case they can be stopped inside the detector.
Such detailed studies are not in the purposes of this work and will be left to further investigation.
The double LHC detection of $\Sigma_f$ decays is open as well at 
very low DM masses, $O(1-100)\mbox{MeV}$ or even lower, in the form of displaced vertices. This kind of scenario, 
as discussed in the previous subsection, can be most likely probed only by collider studies since ID detection is very difficult
\footnote{A possible complementary detection strategy of some classes of FIMP models featuring DM masses of $O(1-100)\mbox{MeV}$ have been actually
proposed in \cite{Essig:2011nj,Essig:2012yx} relying on the direct detection of scattering among the DM and the electrons of detectors like XENON10.
This kind of interaction are not present in hadronic realizations of our framework while in the case of leptonic models the 
scattering cross section results many orders of magnitude below the experimental sensitivity reported in these studies.}.

  \begin{figure}[!htb]
 \begin{center}
  \subfloat{\includegraphics[height=6.5 cm, width=6.5 cm, angle=360]{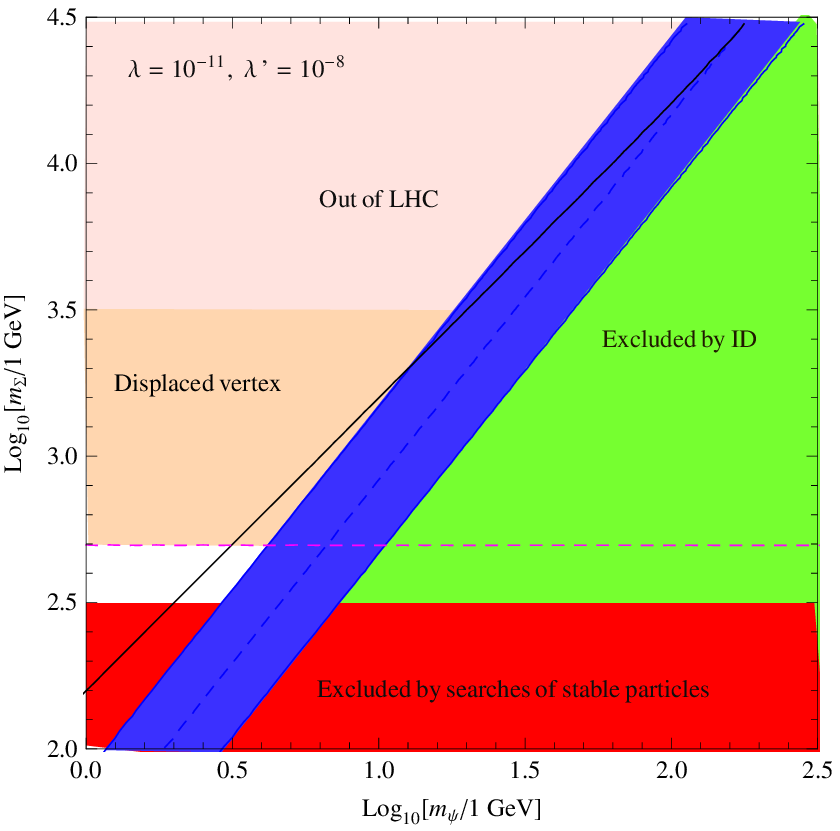}}
  \subfloat{\includegraphics[height=6.5 cm, width=6.5 cm, angle=360]{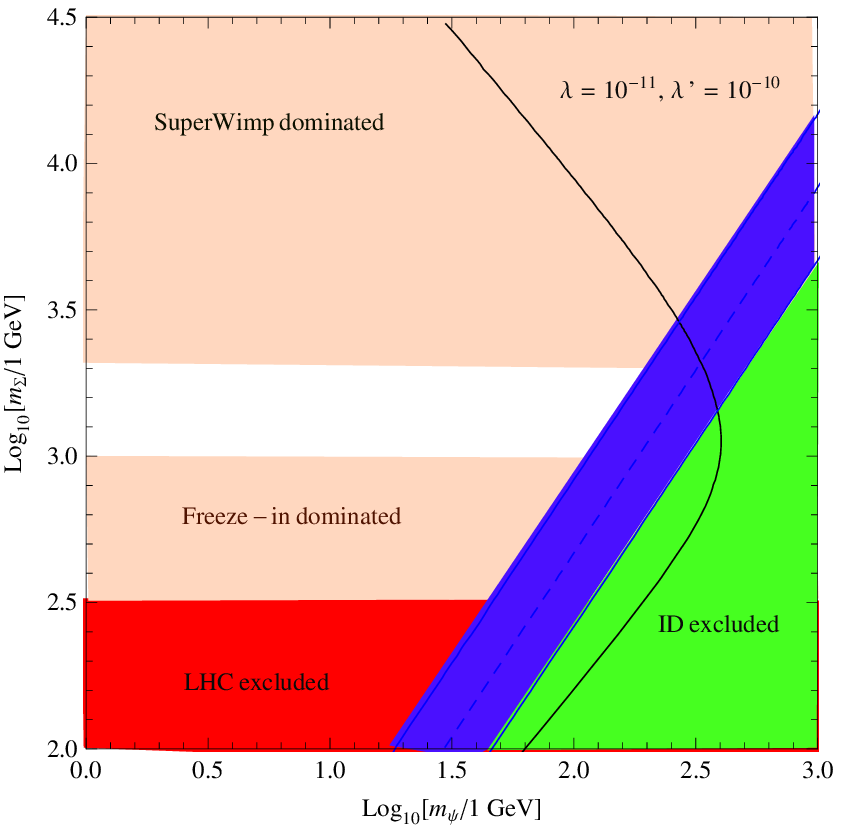}}
 \end{center}
\caption{Summary plot for two models featuring a scalar field electroweak SU(2) doublet. The black lines represent the cosmological value of the DM relic
 density. The blue bands represent an interval of three orders of magnitude of values of the DM lifetime centered around the reference experimental 
 sensitivity of $10^{27} \mbox{s}$. In the model reported on the left panel the field can be detector stable or decay into only SM particles
 through a displaced vertex (until its mass lies
 within the LHC production reach) while in the other case the field can never decay inside the detector. 
 In case of detector stable configurations, the red band
 represents the region exclude by LHC dedicated searches.}
\label{fig:summary_FIMP}  
\end{figure}

In general in the FIMP/SuperWIMP regime, apart the particular scenarios depicted above, 
the only signature accessible at the LHC seems the observation of displaced decays
of pair produced $\Sigma_f$ into only SM states. Such detection would still allow 
to determine the mass and quantum numbers of the scalar state and measure the 
$\lambda'$ coupling. If at the same time also a signal in indirect detection, 
for example in the antiproton flux or in gamma-rays, is detected, the value of 
$\lambda $ could still be inferred from the combined LHC+ID measurements and 
compared with the one required by the FIMP mechanism. On the other hand, also the cosmological 
FIMP scenario can be used to determine 
the missing coupling $\lambda $ and give a prediction of the DM lifetime which can 
be investigated by current and future ID experiments.

Note that for leptonic models, the situation is more complex since the DM relic 
density can be provided also by the SuperWIMP mechanism. Nevertheless in this
case there seem to be a better chance to directly measure the branching ratio
of $\Sigma_f$ at the LHC and determine the coupling $\lambda $, being able
to disentagle the FIMP/SuperWIMP regimes.
We show a summary of two leptonic models in fig.~(\ref{fig:summary_FIMP}),
displaying the range of DM and scalar field masses which can be probed by current and 
future experiments. The blue bands represent the interval $10^{26 \div 28}\mbox{s}$ of 
values of the DM lifetimes. Lower values can be regarded as excluded by ID while 
the regions too far away from these bands cannot account for sizable signals in the 
next future. The black solid lines give the contours of the cosmologically favored 
value of the DM relic density. In the left panel this is achieved through FIMP mechanism. 
The chosen values of the couplings allow the possibility of a detectable signal 
of DM decay and the observation of displaced vertices originated by the 
decay of $\Sigma_f$ into only SM fields as well as a metastable state although, 
in this last case, a sizable portion of the parameter space is already excluded 
by the current LHC searches.
In the right panel the couplings allow both for FIMP and SuperWIMP production,
but the FIMP branch of the curve is already excluded by ID bounds.
In this case the low value of $\lambda^{'}$ leaves the detection of stable 
charged particles as the only collider signature. We finally remark that the two  
 generation mechanisms, namely FIMP and SuperWIMP, might be, at least partially, 
discriminated in case of collider detection of the scalar field. Indeed in case of
 successful reconstruction of its mass and gauge couplings it is possible to determine 
its relic density; a sizable SuperWIMP production can be achieved only if this is 
a least of the order of the thermal value of the DM relic density.

\begin{figure}[!htb]
 \begin{center}
\subfloat{\includegraphics[height= 6.5 cm, width=6.5 cm, angle=360]{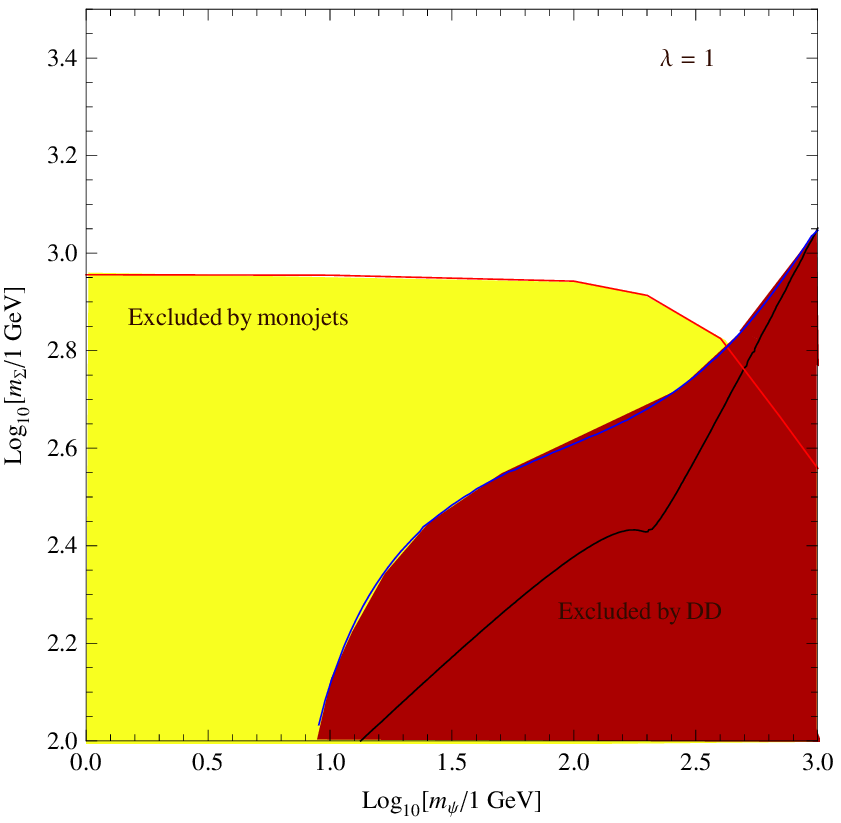}}
\hspace{10 mm}
\subfloat{\includegraphics[height = 6.5 cm, width= 6.5 cm, angle=360]{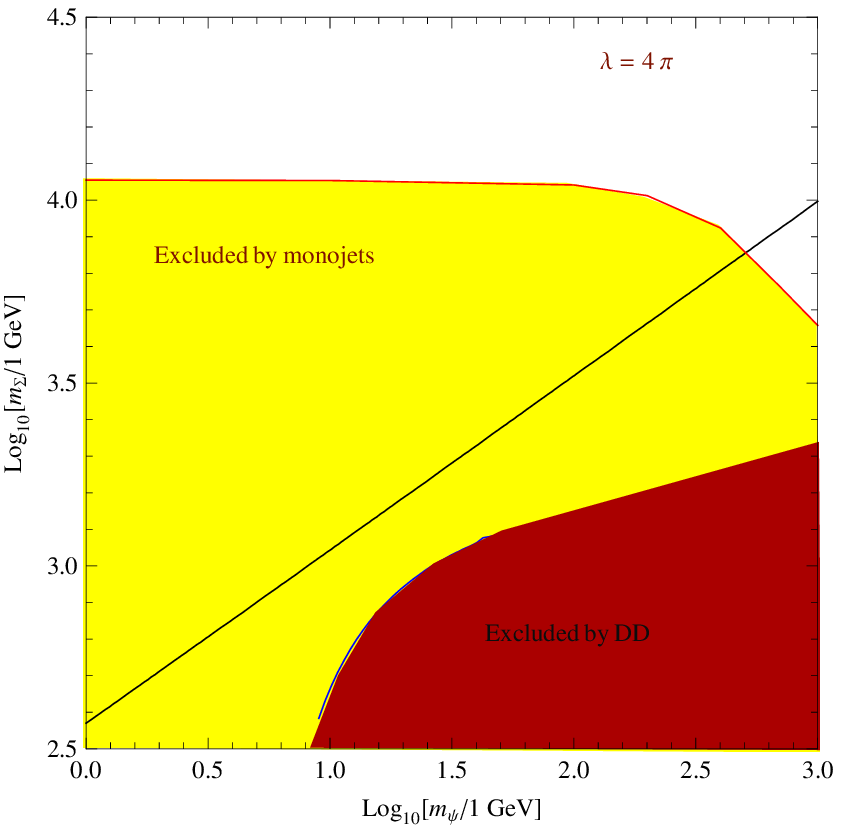}}
 \end{center}
\caption{Summary of the bounds applied in the case of WIMP DM candidate interacting with a colored scalar with coupling $\lambda=1$. The black line represents the cosmological value 
of the DM relic density.The red region is exclude by DM direct detection. The yellow region is instead ruled out by LHC
 monojet searches.}
 \label{fig:summary_WIMP}
\end{figure}

A second result is the investigation of the impact of most recent DD and collider limits 
on a scenario in which the DM candidate, coupled with quarks, behaves like a thermal WIMP. 
Indeed the requirement of the correct dark matter relic density fixes the value of 
$\lambda$ above one, for masses of the scalar not already excluded by LHC. 
This would however imply a too strong signal in direct detection experiments as well as 
in LHC searches of monojets, already excluding large portions of parameters space within 
the LHC reach. This is shown in fig.~(\ref{fig:summary_WIMP}) where we have considered two 
sample models featuring $\lambda=1$ (left panel) and the extremal value $\lambda=4 \pi$, 
the limit value of perturbative couplings. The black curves represent the cosmologically 
favored value of the DM relic density. 
The red colored portion of the parameter space is excluded by DD while the yellow region 
is excluded by monojet searches, which represent the most stringent collider limit. 
In the first case a WIMP dark candidate results entirely ruled out. A viable region of 
parameter space appears by further increasing the coupling $\lambda$. Indeed the requirement 
of the correct relic density requires a sensitive increase of the mass of 
$\Sigma_f$ (cfr. (\ref{eq:lambda_WIMP})). Direct detection  limits result weakened by 
increasing $m_{\Sigma_f}$ and then a viable region may appear at DM masses above 200-300 GeV 
when the limit from monojet searches becomes less effective. Note that in the second panel 
of fig.~(\ref{fig:summary_WIMP}) the field $\Sigma_f$ lies beyond the LHC reach and then 
the only collider probe relies on monojet searches. 

In order to evade these limits one should assume preferred couplings with only third generation 
quarks. In this case the typical signature of this scenario would be the observation direct 
pair production of $\Sigma_{q,u,d}$ with subsequent decay into top or bottom plus missing energy,
translating into events with multiple b-jets, combined with the observation of a signal in 
antiprotons originated by primary products containing as well multiple $b$ or $t$. 
Analogously, leptonic models suffer rather weak limits for monojets and DD and collider, 
with detection in the latter case relying on production and decays of $\Sigma_f$. 
Among the various possibilities the most peculiar is the one featuring the $SU(2)$ 
doublet $\Sigma_l$ since it allows for the combination of a multi-lepton signal at LHC and 
an ID signal in antiprotons.

\section{Conclusions}

The combination of several detection strategies is a very powerful tool in the quest for DM. 
In this paper we have considered the case of decaying Dark Matter in a very simple setup 
which is easily manageable thanks to the small number of parameters involved, but at the 
same time encodes features of popular models like the MSSM with R-parity breaking,
as long as only part of spectrum is light. The model contains only two new particles, 
a Majorana fermion as DM and a scalar field charged under part of the SM gauge group,
coupled to the DM and SM fermions via two different couplings.
 
Our primary goal has been to study the model parameter space accessible both from 
collider searches and current and future ID observations in the regions with a 
cosmologically viable Dark Matter density and highlight the complementary signals. 
From the cosmological point of view, the model is simple since it has a single coupling 
of Dark Matter to the SM and this restricts the possible DM production mechanism. 
Very promising is the FIMP/SuperWIMP regime for low values of the couplings, which 
best fit limits from DM indirect detection. Even such low values of the couplings can 
account for a very rich and not fully explored phenomenology also at LHC. 
In this paper we have identified the most promising regions of the parameter space 
to which we will dedicate a more detailed study of detector (collider as well ID) responses.
While the measurement of both model couplings at the LHC seems to be possible only 
in a very limited region, in most of the parameter space at least the coupling of the additional 
scalar to SM fields is accessible and with the determination of the scalar mass and quantum
numbers allows to predict the ID signal once the second coupling is fixed
to give the right FIMP production. The SuperWIMP mechanism points instead to a heavy metastable scalar 
mostly outside the LHC range, while on the other
hand, if DM is heavy enough, the DM decay could be accessible in ID at CTA and/or AMS-02.
If only the coupling of the scalar $\Sigma_f $ with SM states will be measured at the LHC,
it will be difficult to disentangle this model from supersymmetry with R-parity breaking,
with or without gravitino DM, and an ID signal will be necessary to pinpoint the
identity of Dark Matter. We remark, anyway, that the LHC reach of the class of models 
considered in this work can be potentially enlarged once a quantitative study of the observability of 
long-lived $\Sigma_f$ is performed.

For completeness we have as well considered the case in which one of the couplings 
between the DM and ordinary matter is very large in order to allow for 
the WIMP mechanism. We have investigated whether this scenario is still viable in 
light of the very strong, especially collider, bounds present in particular 
in the case of coupling with quarks. This scenario appears highly disfavored, 
for values of the couplings up to the perturbative limit, unless couplings with 
only third generation quarks are assumed.

\acknowledgments

The authors thank Stephen M. West and Bryan Zaldivar for the fruitful discussions.

The authors acknowledge partial support from the European Union FP7 ITN-INVISIBLES (Marie Curie Actions, PITN-GA-2011-289442).

\appendix

\section{Dark Matter relic density in the FIMP/SuperWIMP regime}

In this section we will describe how our numerical/analytical determinations of the DM relic density in the freeze-in/SuperWIMP regime have been derived. 
On general grounds, our particle physics framework can be described though a system of coupled Boltzmann equations for the $\Sigma_f$ and $\psi$ 
number densities. The dark matter is assumed to never get into thermal equilibrium; we thus assume null initial number density and drop from the equations 
all the terms proportional to this quantity. Under this assumption the relevant equations are given by:
\begin{align}
\label{eq:sigma}
&\frac{dY_{\Sigma}}{d\omega}=-\left(\langle \Gamma \rangle (1-BR) (Y_\Sigma-Y_{\Sigma, \rm eq})-BR \langle \Gamma \rangle Y_\Sigma\right) \frac{1}{H\, \omega} \nonumber\\
&+ {\langle \sigma v \rangle}_{\psi \Sigma} Y_{\Sigma, \rm eq} Y_{\psi, \rm eq} \frac{s }{H\, \omega} \nonumber\\
&- {\langle \sigma v \rangle}_{\Sigma \Sigma} \left(Y_{\Sigma}^2 - Y_{\Sigma, \rm eq}^2\right) \frac{s}{H\, \omega}
\end{align}
\begin{equation}
\label{eq:psi}
\frac{dY_{\psi}}{d\omega}=BR \langle \Gamma \rangle \frac{1}{H\, \omega} + {\langle \sigma v \rangle}_{\psi \Sigma} Y_{\Sigma, \rm eq} Y_{\psi, \rm eq} \frac{s }{H\, \omega}
\end{equation}
where we have adopted the yields $Y=n/s$ and $\omega=m_{\Sigma_f}/T$ as, respectively, dependent and independent variables. $s$ and $H$ are respectively the 
Universe entropy density and the Hubble parameter. The first row of (\ref{eq:sigma}) describes the ${\Sigma_f}$ decays with $BR$ representing the branching ratio into DM. 

In addition to pair annihilation mediated by gauge interactions, with thermally averaged cross section ${\langle \sigma v \rangle}_{\Sigma \Sigma}$, 
the field $\Sigma_f$ can be produced in association with a DM particle by $qg$ or $\gamma l$ annihilations. 
Regarding the DM, it can be produced from $\Sigma_f$ decays or from annihilations of SM states,
in pairs or in association with a $\Sigma_f$ field. 
Although possible, pair annihilation processes of $\Sigma_f$ into two DM particles are not taken into account since the interaction rate is suppressed by $\lambda^4$.

This system of differential equations have been solved through a numerical code built on purpose. 
Due to the relatively strong gauge interactions, the equation for $\Sigma_f$ is dominated by the last row. 
Hence the system can be decoupled and the DM relic density can be computed by solving (\ref{eq:psi}) 
substituting $Y_\Sigma$ with $Y_{\Sigma, \rm eq}$. The equation for the DM can be then cast in the form:
\begin{equation}
\label{eq:analytic_psi}
\frac{dY_\psi}{d\omega}= \Gamma(\Sigma \rightarrow f \psi) 
\frac{g_\Sigma \omega^3 M_{\rm Pl}}{2 \pi^2\, 1.66\, \sqrt{g_{*}} m_{\Sigma_f}^2}K_1(\omega)+A(\omega) 
\frac{45 \omega^{3} M_{\rm Pl}}{16 \pi^6 g_{*}^{3/2} 1.66 m_{\Sigma_f}^4}
\end{equation}
where we have adopted the custom expressions for $H$ and $s$:
\begin{equation}
H=1.66 \sqrt{g^{\rho}_{*}}\frac{T^2}{M_{\rm Pl}},\,\,\,s=\frac{2\pi^2}{45}g_{*}^s T^3
\end{equation}
while $K_1$ is the modified Bessel function. $A(\omega)$ is instead defined as $A=(8\pi^4/T)\,\tilde{A}$ where:
\begin{equation}
\tilde{A} \equiv {\langle \sigma v \rangle}_{\Sigma \psi} n_{\Sigma,\rm eq}n_{\psi, \rm eq}
\end{equation}
This quantity can be expressed according to the general formalism discussed in \cite{Edsjo:1997bg} as: 
\begin{equation}
\tilde{A}(T)=\frac{T}{8 \pi^4}\int_{{\left(m_\psi\right)}^2}^{\infty} g_\Sigma g_\psi K_1\left(\sqrt{s}/T\right) p_1(s) w(s) ds\,.
\end{equation}
where the quantity $w(s)$ is related to the $\Sigma_f \psi $ annihilation cross-section by:
\begin{equation}
 \sigma(s)=\frac{w(s)}{s^{1/2}}p_1(s)
\end{equation}
and can be expressed as \cite{Ellis:1999mm}:
\begin{align}
w(s)=&\frac{1}{4} \int \frac{d^3 p_3}{{\left(2 \pi\right)}^3 E_3} \frac{d^3 p_4}{{\left(2 \pi\right)}^3 E_4} {\left(2 \pi\right)}^4 \delta^4 \left(p_1+p_2-p_3-p_4\right)|M|^2 \nonumber\\
& = \frac{1}{32\pi} \frac{p_3(s)}{s^{1/2}}\int_{-1}^{1} d\cos\theta_{\rm CM} |M|^2
\end{align}
with $| M |^2$ being the invariant squared amplitude (summed over the final state degrees of freedom and mediate over the initial state degrees of freedom) 
and:
\begin{equation}
p_1(s)=p_2(s)=\frac{s-m_{\Sigma_f}}{2 \sqrt{s}}
\end{equation}
\begin{equation}
p_3(s)=p_4(s)=\frac{\sqrt{s}}{2}\,.
\end{equation}
In the case under consideration, namely of $\Sigma_e$, the dominant contribution is, 
as already mentioned, given by the process $\chi \Sigma_e \rightarrow \gamma\,l$,
where $l$ is a charged lepton. The invariant amplitude originates from an s-channel, 
lepton mediated, and a t-channel, $\Sigma_e$ mediated, contribution. 
This quantity can be computed analogously as done in \cite{Ellis:1999mm} 
\footnote{In the computation we have neglected all the masses apart $m_{\Sigma_f}$.}
in the SUSY case leading to:
\begin{equation}
p_1(s)\,w(s)=\lambda^2 \alpha_{\rm em} \frac{-7 m_{\Sigma_e}^4+6 m_{\Sigma_e}^2 s+s^2+4m_{\Sigma_e}^2
\left( s + m_{\Sigma_e}^2\right)\log\left[\frac{m_{\Sigma_e}}{s}\right]}{4 s^{3/2}}
\end{equation}
 Changing the integration variable to $z=\sqrt{s}/T$ we get:
\begin{equation}
A(\omega)= g_\Sigma g_\psi {\left(\frac{m_{\Sigma_e}}{\omega}\right)}^3 \lambda^2 \alpha_{\rm em} \int_{\omega}^{\infty} dz K_1(z) \frac{z^4 + 6 z^2 \omega^2 -7 \omega^4+4 \omega^2 ( z^2 + \omega^2) \log\left[\frac{\omega^2}{z^2}\right]}{4 z^2}
\end{equation}
The DM relic density can be computed by integrating with respect to $\omega$ the right-hand side of (\ref{eq:analytic_psi}) . 
It can be immediately noticed that the scattering and decay contributions have the same dependence on $\lambda$ and
$x$. It is then useful to cast the DM relic density in the form:
\begin{equation}
\Omega h^2 =\frac{m_\psi Y_\psi \left(\omega_0\right)}{3.6\times 10^{-9} \mbox{GeV}}=g_\Sigma \lambda^2 x \left(C_{\rm decay}+C_{\rm scattering}\right)
\end{equation}
where
\begin{align}
& C_{\rm decay}=\frac{135}{1.66\,64 \pi^4} \left(\frac{ M_{\rm Pl}}{1\mbox{GeV}}\right) \times 10^{-3} {\left(\frac{g_{*}}{100}\right)}^{-3/2} \approx 3.4 \times 10^{3} \left(\frac{ M_{\rm Pl}}{1\mbox{GeV}}\right) {\left(\frac{g_{*}}{100}\right)}^{-3/2} \\
& C_{\rm scat}=\frac{45 \alpha_{\rm ew} }{1.66\, 8 \pi^6}\left(\frac{ M_{\rm Pl}}{1\mbox{GeV}}\right) \times 10^{-3}  {\left(\frac{g_{*}}{100}\right)}^{-3/2} \approx 1.3 \left(\frac{ M_{\rm Pl}}{1\mbox{GeV}}\right) {\left(\frac{g_{*}}{100}\right)}^{-3/2} 
\end{align}
where we have defined:
\begin{equation}
I=\int_{0}^{\infty} d\omega \int_{\omega}^{\infty} dz K_1(z) \frac{z^4 + 6 z^2 \omega^2 +-7 \omega^4+4 \omega^2 ( z^2 + \omega^2) \log\left[\frac{\omega^2}{z^2}\right]}{4 z^2}
 \approx 0.46\,.
\end{equation}
From these expressions it is evident that $2 \rightarrow 2$ scatterings give a negligible contribution 
to DM freeze-in in this model.
In the case of a colored scalar the scattering contribution is mainly determined by the process $\Sigma_{f=q,u,d} \psi \rightarrow q g$. The invariant 
amplitude is determined, similarly to the case just depicted, by the the sum of a s-channel and t-channel contribution \cite{Ellis:2001nx}. 
The computation can be performed along the same lines leading to an analogous result as the $\Sigma_e$ case but with an enhancement factor of 
around $\alpha_{\rm s} Q_q/\alpha_{\rm em}$, where $Q_q$ is the charge of the involved quark in 
units of $e$. Even in this case the scattering contribution is at most of the order of percent.

\bibliography{note}{}
\bibliographystyle{hieeetr}

\end{document}